\documentclass[12pt]{iopart}

\usepackage{iopams}
\usepackage{amscd}
\usepackage{graphicx}
\usepackage{bm}
\usepackage{color}
\usepackage{xcolor}
\usepackage{times}
\usepackage[normal]{subfigure}
\usepackage{float}
\usepackage{setspace}
\usepackage{dcolumn}
\usepackage{epstopdf,epsfig}

\begin{document}

\title{Quantum Otto cycle with inner friction: finite time and disorder effects}

\author{A. Alecce,$^{1}$ F. Galve,$^{2}$ N. Lo Gullo,$^{1,3}$ L. Dell'Anna,$^{1,3}$ F. Plastina,$^{4,5}$ R. Zambrini$^{2}$}

\address{
 {1} Dipartimento di Fisica ed Astronomia G. Galilei,  Universit\`a di Padova, Via
Marzolo 8 Padova (PD) Italy\\
     {2} IFISC (UIB-CSIC), Instituto de F\'isica Interdisciplinar y Sistemas Complejos, UIB Campus,
E-07122 Palma de Mallorca, Spain\\
 {3} CNISM, Sezione di Padova, Italy\\
   {4}  Dip.  Fisica, Universit\`a della Calabria, 87036
Arcavacata di Rende (CS), Italy  \\
   {5} INFN - Gruppo collegato di Cosenza, Cosenza Italy}
\ead{roberta@ifisc.uib-csic.es}
\vspace{10pt}
\begin{indented}
\item[]March 2015
\end{indented}

\begin{abstract}
The concept of inner friction, by which a quantum heat engine is
unable to follow adiabatically its strokes and thus dissipates
useful energy, is illustrated in an exact physical model where the
working substance consists of an ensemble of misaligned spins
interacting with a magnetic field and performing the Otto cycle.
The effect of this static disorder under a finite-time cycle gives
a new perspective of the concept of inner friction under realistic
settings. We investigate the efficiency and power of this engine
and relate its performance to the amount of friction from
misalignment and to the temperature difference between heat baths.
Finally we propose an alternative experimental implementation of
the cycle where the spin is encoded in the degree of polarization
of photons.

\end{abstract}

%
%
%
\maketitle
%
%

\section{Introduction}
The recent boosting interest in the study of the quantum
counterpart of classical well-known heat engines such as Otto,
Carnot, Stirling and Szilard
ones~\cite{cycles,otto1,otto2,carnot1,carnot2,stirling1
,szilard,campisi} has been motivated both
 by the need of a fundamental understanding
of the limits imposed by quantum mechanics on the thermodynamic
performances of small devices (in terms of both efficiency and
power output), and by the growing experimental ability to control
various types of quantum systems with a high degree of accuracy.
There have been, indeed, many proposals aimed at implementing
thermodynamic transformations and cycles with many different
quantum working substances, ranging from trapped ions to magnetic
materials \cite{azimi}, with the prospect of building quantum heat
engines, exploring the abilities and limitations of quantum
machines in converting heat into work, and, on more general
ground, to build a self contained description of thermodynamics in
the quantum regime. As a specific example, an interesting proposal
in this respect have been made for implementing a nanoheat engine
with a single trapped ion, performing a quantum Otto cycle
\cite{lutztrapped}. Besides its specific applications, this is an
important example as the quantum Otto cycle constitutes a useful
test ground to study irreversibility in the quantum realm.

Indeed, the cycle consists of two isochoric thermalization
branches (with a fixed system Hamiltonian) and two isentropic
branches in which the system is detached from the thermal baths
and its evolution is generated by a parametric time-dependent
Hamiltonian. Every practical realization of these latter
transformations has to face the general problem of understanding
and describing the (un-wanted) irreversible entropy production
which 
can occur in non-ideal, finite time quantum parametric processes.
This general problem has been variously analyzed through the use
of fluctuation relations \cite{jarz,innerfriction}, and has
attracted a lot of attention in recent years \cite{rev,meas}.

In this paper, we explicitly address the study of the Otto cycle
by focusing on the finite time case and discussing the
implications of finite time transformations as opposed to ideal
infinitely lasting ones. In this respect, in a series of papers,
Kosloff and Feldmann~\cite{kosloff2000,kosloff1,kosloffMANY}
introduced the concept of intrinsic/inner friction, whereby the
engine is never able to accomplish a frictionless adiabatic
transformation and thus looses power. This concept has been then
extended and applied to various contexts
\cite{innerfriction,wang12,wang}.

Inner friction is a fully quantum phenomenon, whose consequences
are similar to those of the mechanical friction occurring when
displacing a piston in compressing/expanding a gas in a classical
thermodynamic setting. Its origin, however,, is completely
different: when the external control Hamiltonian does not commute
with the internal one, the states of the working fluid cannot
follow the instantaneous energy levels, leading to additional
energy stored in the working medium. Inner friction is thus
associated to diabatic transitions, i.e. changes of populations
which occur during the time dependent adiabatic (here referring to
closed system) strokes if they are performed at finite speed.

So far inner friction occurring in specific cycles and
transformations has been analyzed by adopting phenomenological and
physically motivated assumptions about the explicit time
dependence of the population changes (e.g., in Ref.
\cite{kosloff2000}, a friction coefficient is introduced, giving
rise to a constant dissipated power). Our treatment, instead, does
not rely on any ad hoc assumption, but rather on the exact
dynamics of the working substance. This is important because it
has been shown, \cite{innerfriction}, that inner friction is not
only an indicator of irreversibility of a quantum process, but
also a quantitative measure of its amount. It is therefore crucial
to identify and highlight its role in the efficiency reduction of
finite time cycles by analyzing the full quantum dynamics that
produces it.

In particular, we will explore the quantum friction arising from
disorder within the sample playing the role of working substance.
We will consider an ensemble of qubits in a setting in which their
Hamiltonian parameters are not homogeneous, and connect the
presence of these {\it static errors} to the appearance of
friction and losses during the implementation of the Otto cycle.
Explicitly, we provide a {\it quantitative} analysis of the amount
of losses due to the inner friction as a function of the degree of
disorder.

Indeed, the performance of the heat machine are negatively
affected by inner friction and cycle's outputs such as extracted
work, power and efficiency are gradually suppressed as disorder
and friction increase.

The remainder of the paper organized as follows. In
Secs.~\ref{sec:friction} and \ref{sec:ancora} we introduce and
review the concept of inner friction by focusing on the particular
case of a spin system in presence of misalignments and disorder,
which will then be of interest for the rest of the present paper.
In Sec.~\ref{sec:otto} we introduce the quantum Otto cycle and its
constituent transformations specifying the assumptions about the
model we use to describe the working substance. In
Sec.~\ref{sec:results} we present and discuss our main results,
while in Sec.~\ref{sec:exp} we propose a feasible experimental
implementation of the quantum Otto cycle in order to test our
findings. Finally, Sec.~\ref{sec:conclusions} is devoted to some
concluding remarks and to a discussion of possible future
developments.

\section{Model and methodology}\label{sec:model}

In this section we introduce the model, and give a possible
explanation of the origin of inner friction. We then introduce the
quantum Otto cycle (QOC) and the figures of merit through which
the cycle will be characterized.

\subsection{Misalignment and disordered samples}
\label{sec:friction}

In order to understand what we mean by losses and friction in a closed quantum system, and
in particular in the case of one qubit, let us focus on the dynamics generated by a Hamiltonian of the form:
\begin{equation}
\label{eq:adiabham}
H(\lambda(t))=\frac{\omega_0}{2}\sigma_z+\lambda(t)(\cos\theta\sigma_z+\sin\theta\sigma_x).
\end{equation}
The analysis reported here applies to the general case of qubit
dynamics (\ref{eq:adiabham}) and in the following we will consider
the case of a spin interacting with a magnetic field (Sections 2
and 3) as well as of a qubit encoded in photon polarization
(optical implementation in Section 4).

An adiabatic transformation is obtained by the unitary time
evolution generated by the Hamiltonian (\ref{eq:adiabham}), with a linear driving  of
the external field at a fixed rate $\lambda (t)=\alpha \omega_0 t/2$,
which we allow to be misaligned by an angle $\theta$ with respect
to the static field $\omega_0$. The misalignment affects
the energy spacing as well as the eigenstates and the populations.

We assume that at $t=0$ the qubit in a thermal state at inverse
temperature $\beta$. For a very slow driving, ideally taking an
infinite time to complete the transformation in the quantum
adiabatic regime, the qubit populations would remain unchanged
while the  the energy spacing increases/decreases and the system
remains in a thermal state with a lower/higher temperature. The
same occurs in absence of misalignment, $\theta=0$ in Eq.
(\ref{eq:adiabham}), as in this simple scenario the adiabatic
transformation
\begin{equation}
\label{eq:zedev} H=\omega(t)\sigma_z \, , \quad \mbox{with }
\omega(t) = \frac{\omega_0}{2} + \lambda(t)
\end{equation}
reduces to a compression/expansion of the energy spacing of the qubit,
thus preserving the initial thermal populations  even in presence
of fast driving.

\begin{figure}
\centering
\includegraphics[width=.45\columnwidth]{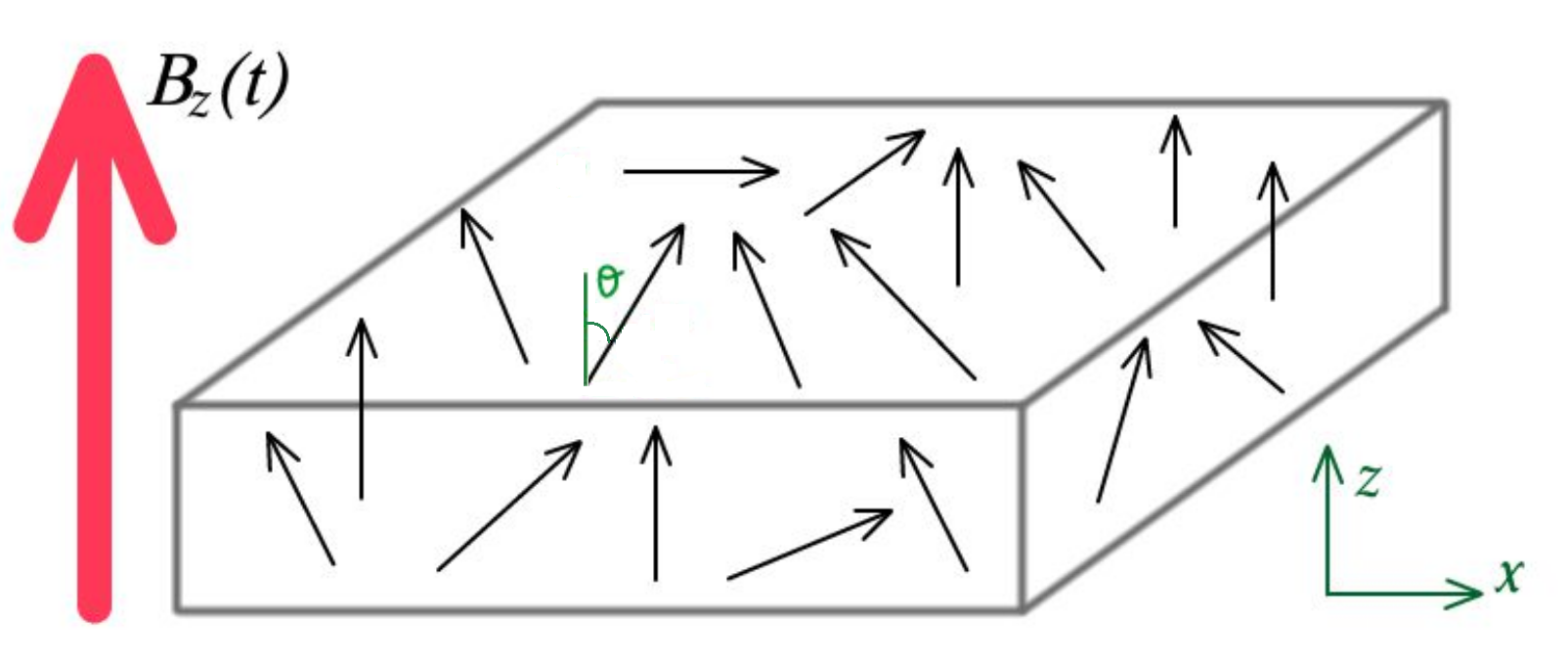}(a)~~~\includegraphics[width=.4\columnwidth]{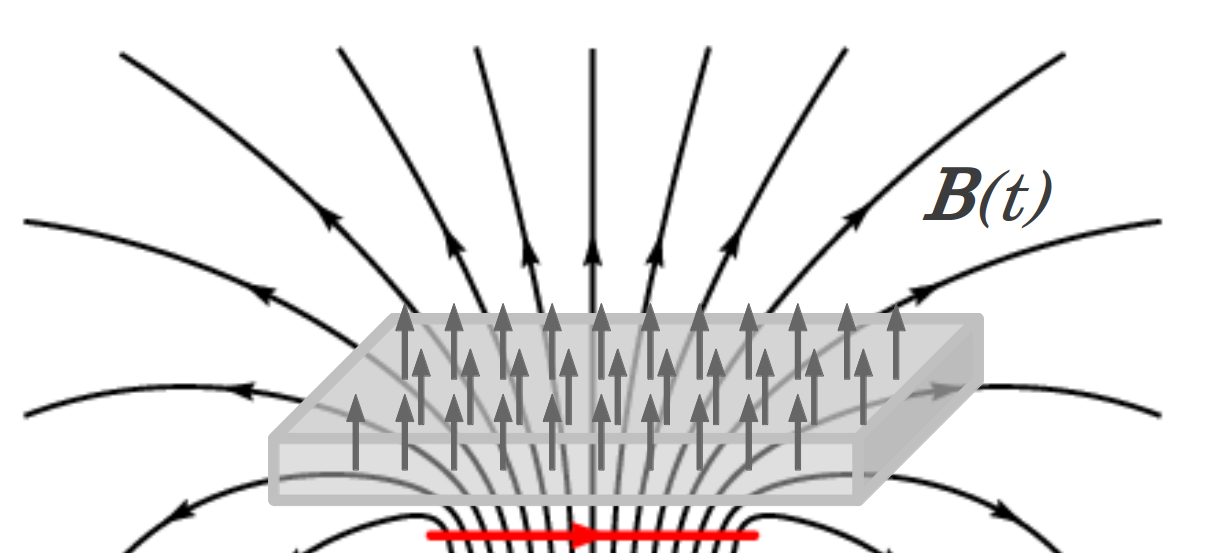}(b)
\caption{Misalignment effects of a sample of spins interacting
with an external field due to (a) disorder within the sample or
(b) lack of uniformity of the magnetic field (e.g. because of the
finite length of the coil).} \label{fig:scheme}
\end{figure}

Interesting dynamical and thermodynamical implications arise,
instead, when considering deviations from the limit of perfect
alignment ($\theta\neq 0$). This is the case that we are going to
explore in this work in order to characterize inner friction and
its effects on the efficiency of quantum thermal machines,
resulting from the simultaneous presence of the static field
$\omega_0$ and of the misaligned time depending part $\lambda(t)$.
Our aim it to apply this analysis to an ensemble of (independent)
spins, considering some degree of disorder and looking at average
effects across the sample. In particular, this can correspond to
different situations as represented in \Fref{fig:scheme}. A
condensed system on a lattice, with embedded magnetic dipoles
having disordered orientations, can be modelled by randomly
oriented spins with tilting angles $\theta_i$ ($i=1,2, \ldots$)
with respect to the direction of a uniform external field. We
assume that the distribution of the spin orientations in the
sample is given by a function $G(\theta)$. Alternatively, all
sample dipoles could be perfectly aligned in an ordered
configuration but the inner friction could be due to inhomogeneity
of the external fields in space 
(\Fref{fig:scheme}b). The field orientation across the sample
would be given, in this case, again by the function $G(\theta)$.

\subsection{Inner friction and irreversibility}
\label{sec:ancora}

In order to have a simple physical picture for the behavior of our
quantum machine, let us first consider the simpler case of a
driven quantum two level system undergoing the unitary dynamics
generated by a parametric time dependent Hamiltonian
$H[\lambda(t)]$. If the parameter $\lambda(t)$ changes slowly
enough (in the sense of the quantum adiabatic theorem,
\cite{messiah,armin}) the system evolves without its energy
population ever changing at all, even if the instantaneous energy
eigenvalues and eigenstates do change in time. If the system has
been prepared in equilibrium with a thermal bath, which is then
removed, such an ideal adiabatic parameter change keeps the system
in an equilibrium state at every stage.  In particular, if the
parameter $\lambda$ gets back to its initial value after some
time, the final result is that the system is brought back to its
initial state. On the other hand, if the cycle is performed in
finite time, the final state of the system will differ from the
equilibrium state it started off because non-adiabatic transition
have taken place \cite{nonad}. The difference between the two
states, if properly quantified, can be regarded as a measure of
the deviation from an ideal adiabatic transformation. The quantum
non-adiabaticity has the same effects as friction has in a
classical context: an extra energy is needed to complete the
process (indeed, the work done in the ideal adiabatic is always
smaller than the one performed in finite time, see
\cite{armin,innerfriction}), which is then dissipated if the
system equilibrates at the end of the process.

With this picture in mind, let us now address the dynamics
generated by Hamiltonian (\ref{eq:adiabham}) on an initial thermal
state given by $\rho_0=\exp\{-\beta
H(\lambda(0))\}/\Tr[\exp\{-\beta H(\lambda(0))\}]$ where $\beta$
is the inverse temperature in units of the Boltzmann constant. By
changing $\lambda(t)$ very slowly from $\lambda(0)$ at $t=0$ up to
$\lambda(t_f)=\lambda^*$ at $t=t_f$ and then going back from
$\lambda^*$ to $\lambda(0)$ the system will be brought back into
its initial state. To discuss what happens in the general case,
namely when these changes are performed at finite rates, we
consider the following protocols:

\begin{equation}
\begin{CD}
\rho_{0} @>U_F(0,t_{F})>> \rho_1 \\
\rho_2@<< U_B(0,t_{B})< \rho_1 \\
\end{CD}
\end{equation}

The forward protocol, defined by the unitary operator
$U_F(0,t_{F})=\mathcal{T}e^{-\imath\int_{0}^{t_{F}}
H(\lambda_F(\tau)) d\tau}$ ($\mathcal{T}$ being the time ordering
operator), is generated by the Hamiltonian in
Eq.~(\ref{eq:adiabham}) such that $\lambda_F(t)=\alpha_F\omega_0
t/2$. It takes the initial density matrix $\rho_0$ to
$\rho_1=U_F(0,t_{F})\rho_0 U_F^{\dag}(0,t_{F})$. The backward
protocol $U_B(0,t_{B})=\mathcal{T}e^{-\imath\int_{0}^{t_{B}}
H(\lambda_B(\tau)) d\tau}$ is again generated by the Hamiltonian
in Eq.~(\ref{eq:adiabham}) where now
$\lambda_B(t)=\lambda_F(t_F)-\alpha_B\omega_0 t/2$ with the
condition that $\lambda_B(t_{B})=\lambda_F(0)$. This consists just
in ramping up and down the field $\lambda(t)$ with different rates
$\alpha_F$ and $\alpha_B$, respectively.

\begin{figure}
\centering
\subfigure{
\includegraphics[width=.4\columnwidth]{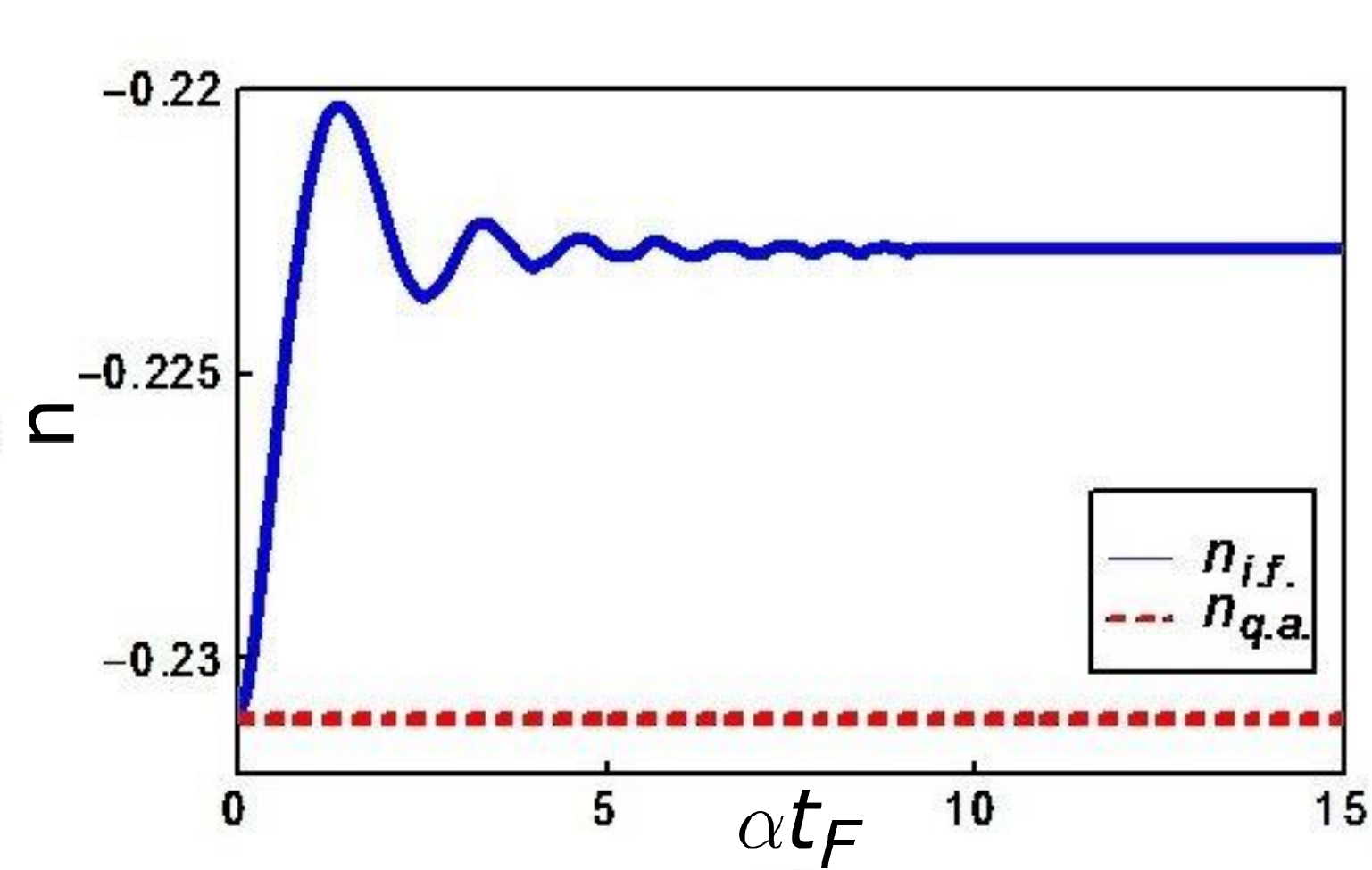}
\label{fig:polarization}} \hspace{2mm}
\subfigure{
\includegraphics[width=.4\columnwidth]{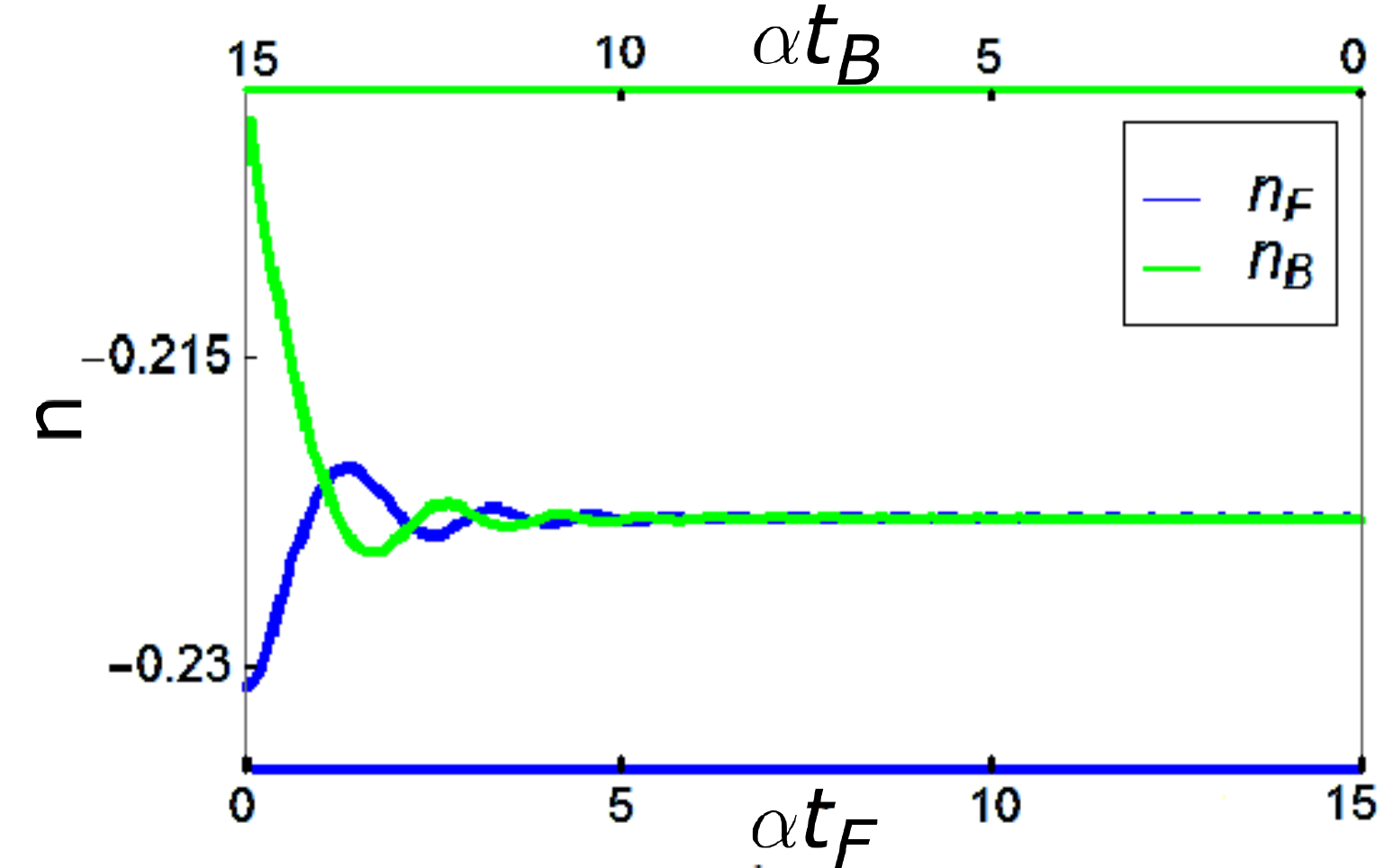}
\label{fig:fvb} } \caption{The left panel shows the time
dependence of the polarization $n(t)$ for a finite-time adiabatic
transformation (solid blue line) compared with the case of an
ideal quantum adiabatic one (dashed red line). We used
$\alpha_F=10^{-4} \omega_0 $. In the right panel,  we compare the
forward evolution (lower time axis from left to right) with the
backward one (upper time axis from right to left) by displaying
the time dependent polarization in both processes. The parameters
used  are $\alpha_F t_{F}=\alpha_B t_{B}=15$ and
$\alpha_F=\alpha_B=\omega_0$. In both figures $\theta=\pi/5.$}
\end{figure}

In order to characterize the above protocol, we first look at the
time dependent polarization defined as
$n(t)=\Tr[\rho(t)H(t)]/\omega(t)$ where $\omega(t)$ is the energy
level spacing at time $t$ for both the forward and backward
protocols. The result is shown in Fig.~\ref{fig:polarization},
where we notice that finite-time evolution introduces deviations
with respect to the quantum adiabatic case as expected. Moreover,
as it can be seen in Fig.~\ref{fig:fvb}, by applying the forward
and backward protocols defined above, the system does not get back
to its initial state, but reaches a different polarization (green
line) at the end of the protocol.
This already gives a quantitative indication that finite time
control leads to an irreversible behavior. Here, we use the word
``irreversibility'' in the thermodynamic sense: because of the
occurrence of non-adiabatic transitions, the system is driven out
of the manifold of equilibrium states and application of the same
protocol in reverse does not bring it back to the initial state.

A more precise way of quantifying the irreversibility of such a
transformation is through the distance of the final state from the
initial one, expressed in terms of the relative entropy
$D(\rho_2||\rho_0)$, where $\rho_2= U_B(0,t_{B})\rho_1
U_B^{\dag}(0,t_{B})$. As shown in \cite{innerfriction}, this
quantity has a well defined thermodynamical interpretation as it
precisely gives the non-adiabatic part of the work performed on
the system by the driving agent, i.e. the inner friction.

Indeed, for an adiabatic transformation, the quantum relative
entropy between the actual final state and the ideal thermal
equilibrium one is proportional to the difference between the work
done on the system during the parametric change and the same
quantity taken in the infinitely slow limit. This is precisely the
definition of the inner friction, hereafter called $W_{fric}$,
\cite{innerfriction}. Furthermore, the same quantity is linked to
the generation of extra heat; that is to say, to the irreversible
production of `waste energy'. This extra energy is exactly the
energy that needs to be dissipated if, at the end of the protocol,
we were to thermalize the system to the initial temperature.
Specifically, the following relations hold:
\begin{equation}
- \beta Q(\rho_2 \rightarrow \rho_0)=\beta W_{fric}=
D(\rho_2||\rho_0) \label{entropy}
\end{equation}
where $Q(\rho_2\rightarrow \rho_0)$ is the heat the system takes
to thermalize at the initial inverse temperature $\beta$. This is
what we shall refer to as inner friction in the following.

\begin{figure}[H]
\centering
\includegraphics[width=9cm]{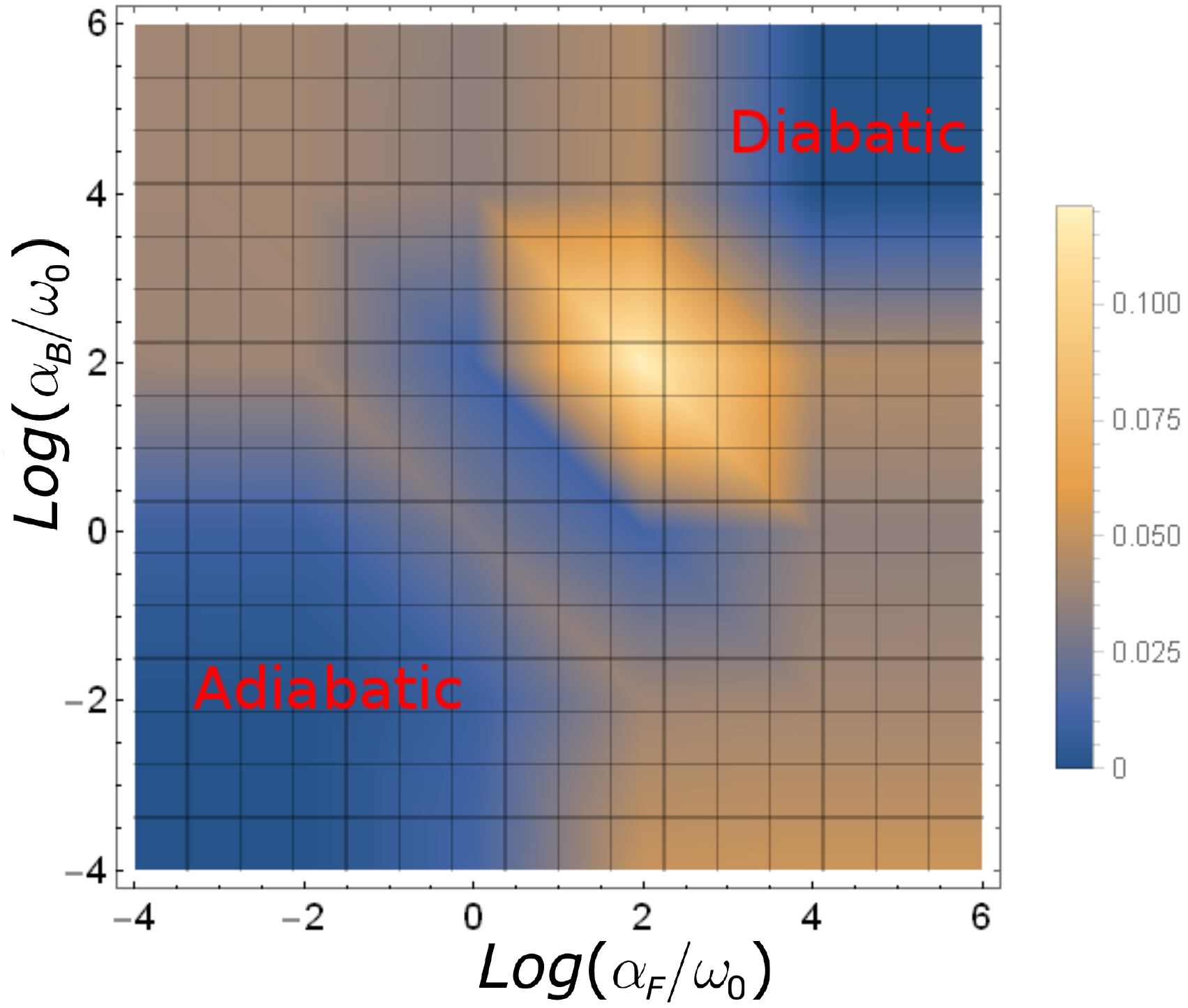}
\caption{The quantum relative entropy between the state at the end
of the backward step and the initial thermal state $\rho_{0}$.
Notice that inner friction is very close to zero both for
$\alpha_{F,B}\rightarrow 0$ and $\alpha_{F,B}\rightarrow\infty$.
The inverse temperature characterizing the initial state is taken
to be equal to the energy spacing $\omega_0$.
Here $\theta=\pi/5$ and $\alpha_F t_{F}=\alpha_B t_{B}=15$.}
\label{fig:relativeentropy}
\end{figure}

The inner friction for the time evolution described above is
reported in Fig.~\ref{fig:relativeentropy}, where we can clearly
see that when both transformations are either very slow (quantum
adiabatic case) or very fast (`diabatic' or sudden case), at the
end of the protocol the system is found to be in (or very close
to) its initial state. For finite time transformations, however,
the system does not get back to its initial state. From a
dynamical point of view, this is not surprising; however, if
interpreted from a thermodynamical perspective, this fact suggests
that transformations done in finite time are, in general,
irreversible ones.

\subsection{Model for the quantum Otto cycle (time scales assumptions)}
\label{sec:otto}

The Otto cycle is the simplest cycle for our purposes as it allows
for a clear separation between dissipative steps (thermalization
processes, in contact with a thermal bath) from non-dissipative
ones (in which work is done or extracted), as opposed, for
instance, to the Carnot cycle, which contains two isotherms in
which one has to perform (extract) work while the system is
attached to a thermal bath. This separation will be very useful in
order to identify finite time effects on the single adiabatic
transformations and thus on the total cycle.

The quantum version of the Otto cycle is the composition of two
adiabatic transformations, in which the systems evolves unitarily,
and two isochoric branches corresponding to thermalization in
contact with a hot (and, respectively, a cold) heat bath at
temperature $\beta^{-1}_h$ ($\beta^{-1}_c$ ).

In the next subsections we better specify the assumptions we
employ to describe the different branches of the quantum Otto
cycle and the physical quantities we investigate to characterize
it. The ideal Otto cycle is represented by the dashed (yellow)
rectangle in Fig. \ref{fig:phasespace}. The blue line, instead,
describes a finite time cycle, in which the end points of the
adiabatic strokes are moved towards larger values of $n$ (which
just means that there is more population than expected in the
excited states) because of the presence of inner friction.

\subsubsection{Adiabatic transformation - }
As already mentioned above, the adiabatic transformations can be
described by the unitary operator generated by the Hamiltonian in
Eq. (\ref{eq:adiabham}). For simplicity, in the following we will
consider the case where the two adiabatic branches ($1\rightarrow 2$
and $3\rightarrow 4$) last equally
long, namely $t_F=t_B=\tau_{ad}$ and have the same rate of change
for the field $\alpha_F=\alpha_B$.

\subsubsection{Isochoric transformations -}
For the isochoric transformations, 
we assume perfect thermalization at the given temperatures
$\beta^{-1}_h$, $\beta^{-1}_c$ (hot and cold, respectively). To
study the relation between inner friction and the finite time of
the adiabatic branches, we will assume that perfect thermalization
is achieved very fast with respect to all other time scales, and
also that the isochoric branch will be assigned with a fixed short
time duration (to be eventually neglected)
with respect to adiabatic ones but long with respect
to thermalization time of the system:
\begin{equation}
\tau_{therm}\ll \tau_{iso}\ll \tau_{ad}
\end{equation}
where $\tau_{therm}$, $\tau_{iso}$ and $\tau_{ad}$ are the typical
time scales for the thermalization process, isochoric and
adiabatic transformations, respectively.

\begin{figure}
\centering
\includegraphics[width=9cm]{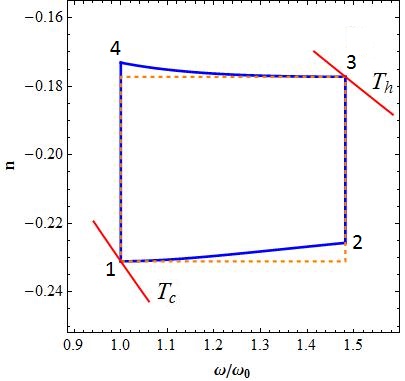}
\caption{Representation of the Otto cycle in a parameter space in
which the horizontal axis gives the instantaneous energy spacing
between the eigenstates of the Hamiltonian (\ref{eq:adiabham}),
while the vertical axis gives the polarization. The solid blue
line is an example of finite-time Otto cycle with parameters:
$\theta=\frac{\pi}{5}$, $\alpha t=\omega_0 t=0.5513$, $\beta_c=
\omega_0^{-1}$ and $\beta_h= \beta_c/2$. 
The dashed yellow  line corresponds to an ideal (infinite time)
Otto cycle. The two red lines are the isotherms in this plane.
They include the two (very fast) branches in which the system
equilibrates in contact with baths at inverse temperatures
$\beta_c$ and $\beta_h$, respectively. \label{fig:phasespace}}
\end{figure}

\subsection{Figures of merit}

In order to characterize the quantum Otto cycle, we will look at
the extractable work, $W_{ex}$, at the power $\mathcal{P}$, at the
efficiency $\eta$, and at their averages over disorder.

To properly define these quantities, let us start by defining the
work done on an adiabatic branch as:
\begin{equation}
 \label{eq:work}
 W=\Tr[H_f\rho_f]-\Tr[H_i\rho_i]
\end{equation}
where $H_i\;(H_f)$ and $\rho_i\;(\rho_f)$ are the Hamiltonian and
the density matrix of the system at the beginning (end) of each
transformation. In particular, both adiabatic
transformations start with a Gibbs-like state since we assume
perfect thermalization to occur at the end of each isochores. In
the adiabatic transformations, the work defined in
Eq.~(\ref{eq:work}) does coincide with the first moment of the
work distribution for closed but non-autonomous systems
\cite{lutzwork}. On the other hand such a work distribution allows
to define a fluctuation relation and thus its moments have a clear
thermodynamical meaning.

In the isochoric branches, we have that the initial and final
Hamiltonians are the same and the final state $\rho_f$ is thermal,
and thus diagonal in energy eigenbasis. The amount of energy
exchanged between the reservoir and the system in each isochoric
transformation is given by:

\begin{equation}
 \label{eq:workiso}
 Q_{iso}=\epsilon_1 \left(p_1^{(f)}-p_1^{(i)}\right) + \epsilon_0 \left(p_0^{(f)}-p_0^{(i)}\right)\equiv \omega \left(p_0^{(i)}-p_0^{(f)}\right).
\end{equation}

Thus, the energy absorbed from the bath equals the energy spacing
$\omega= \epsilon_1-\epsilon_0$, times the change in the
population of the lowest energy state (we denoted the ground and
excited state populations as $p_0$ and $p_1$, respectively).

Since  the change of the total internal energy along the cycle
vanishes, the total work done on the system is given by
$W_{tot}=-(Q_{h}+Q_{c})$ where $Q_{h}$ ($Q_{c}$) is the amount of
energy  exchanged with the reservoir at inverse temperature
$\beta_h$ ($\beta_c$), given by Eq. (\ref{eq:workiso}) for the isochores
$2\rightarrow 3$ ($4\rightarrow 1$). The first quantity we will use to
characterize the cycle is the extractable work
$W_{ex}=-W_{tot}=(Q_{h}+Q_{c})$ given by the relation:

\begin{equation}
 \label{eq:totwork}
 W_{ex}=\left(\omega_{2}(p_0^{(2)}-p_0^{(3)})+\omega_{1}(p_0^{(4)}-p_0^{(1)})\right),
\end{equation}

\noindent where $\omega_{k}=(\epsilon_1^{(k)}-\epsilon_{0}^{(k)})$
is the energy level spacing of the Hamiltonian at point $k=1,2$ in
the $\omega$-$n$ diagram of Fig.~\ref{fig:phasespace}.

For the quantum Otto cycle, and by means of the definition of $n$,
we can then write the following condition

\begin{equation}
 \omega_1(n_{(1)}-n_{(4)})<\omega_2(n_{(2)}-n_{(3)}),
\label{eq:W>0}
\end{equation}
ensuring that the work extracted is strictly positive and we are
actually using the engine to perform work. This is in agreement
with Carnot theorem, as shown in Ref.~\cite{kosloff2000}, and for
ideal quantum adiabatic branches ({\it i.e.} infinitely slow
transformations) it reduces to $\beta_h/\beta_c>\omega_1/\omega_2$
where $\omega_{1,2}$ are the energy spacings at the end of {\it
quantum adiabatic} branches \cite{kieu}. Notice that, in this
quantum adiabatic limit, a breakdown of the positive work
condition coincides with the saturation of Carnot inequality.

Let us now define the power and the efficiency of a cycle on a single qubit as:

\begin{eqnarray}
 \label{eq:power}
& \mathcal{P}(\tau_{ad},\theta,\beta_h/\beta_c)=\frac{W_{ex}}{t_F+t_B+\tau_{iso}}=
\frac{W_{ex}}{2\tau_{ad}+\tau_{iso}},\\
& \eta(\tau_{ad},\theta,\beta_h/\beta_c)=\frac{W_{ex}}{Q_{h}}=1+\frac{Q_c}{Q_h}
\end{eqnarray}
where we have made explicit the dependence upon the angle $\theta$
in both $\mathcal{P}$ and $\eta$ and we explicitly included
$\tau_{iso}$ in the above definitions, despite the assumption that
it is small compared to the characteristic times scale of the
cycle, just to avoid that $\mathcal{P}$ is ill-defined in the
formal limit $\tau_{ad}\rightarrow 0$.

In addressing the disordered case we average all quantities over
$\theta$ assuming it to have a Gaussian distribution
$G_\sigma(\theta)$ with zero mean and variance $\sigma^2$. The
averaged extractable work, power and efficiency are thus given by:

\begin{eqnarray}
\overline{W}_{ex}(\tau_{ad},\beta_h/\beta_c,\sigma)=\int_0^\pi G_\sigma(\theta)W_{ex}(\tau_{ad},\theta,\beta_h/\beta_c)d\theta\\
\overline{\mathcal{P}}(\tau_{ad},\beta_h/\beta_c,\sigma)=\int_0^\pi G_\sigma(\theta)\mathcal{P}(\tau_{ad},\theta,\beta_h/\beta_c)d\theta\\
\overline{\eta}(\tau_{ad},\beta_h/\beta_c,\sigma)=\int_0^\pi G_\sigma(\theta)\eta(\tau_{ad},\theta,\beta_h/\beta_c)d\theta.
\end{eqnarray}

\section{Results and discussion}

\label{sec:results}

In this section we characterize the QOC by looking at the
extractable work $W_{ex}$, its power $\mathcal{P}$ and its
efficiency  $\eta$, paying particular attention to the role of the
inner friction in limiting the performances of such an heat
engine. In the first subsection we will look at these figures of
merit for different sets of parameters of our model-system. In the
second one, we study the behavior of the efficiency at maximum
power, $\eta (\mathcal{P}_{MAX})$.

\begin{figure}
\centering

\subfigure[Extractable work ($W_{ex}$) vs. $\alpha t_{tot}$ for
different values of $\theta$] {
\includegraphics[height=4cm,width=.42\columnwidth]{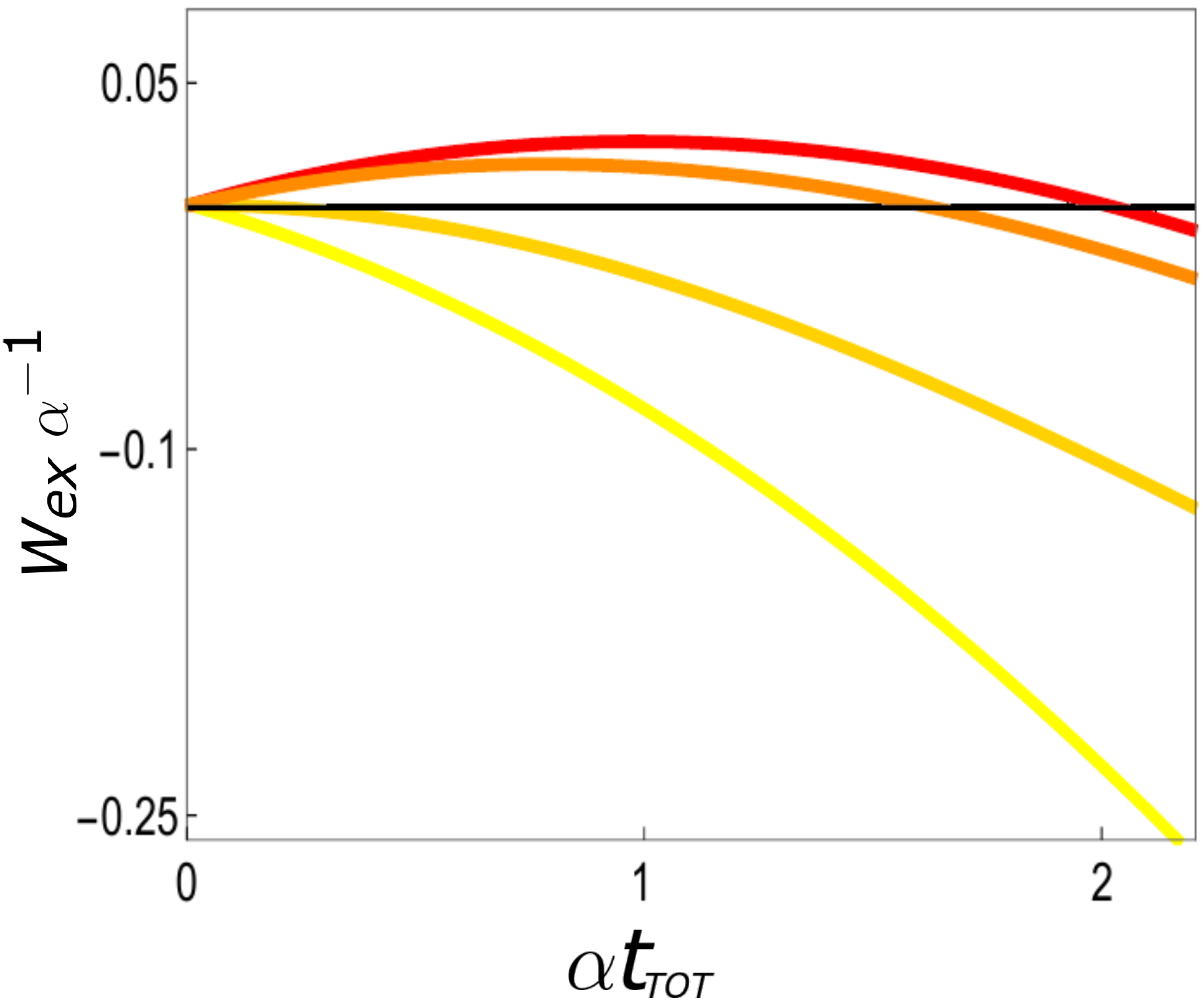}
\label{fig:W-t(theta)}} \subfigure[Extractable work ($W_{ex}$) vs.
$\alpha t_{tot}$ for various $\beta_h/\beta_c$] {
\includegraphics[height=4cm,width=.42\columnwidth]{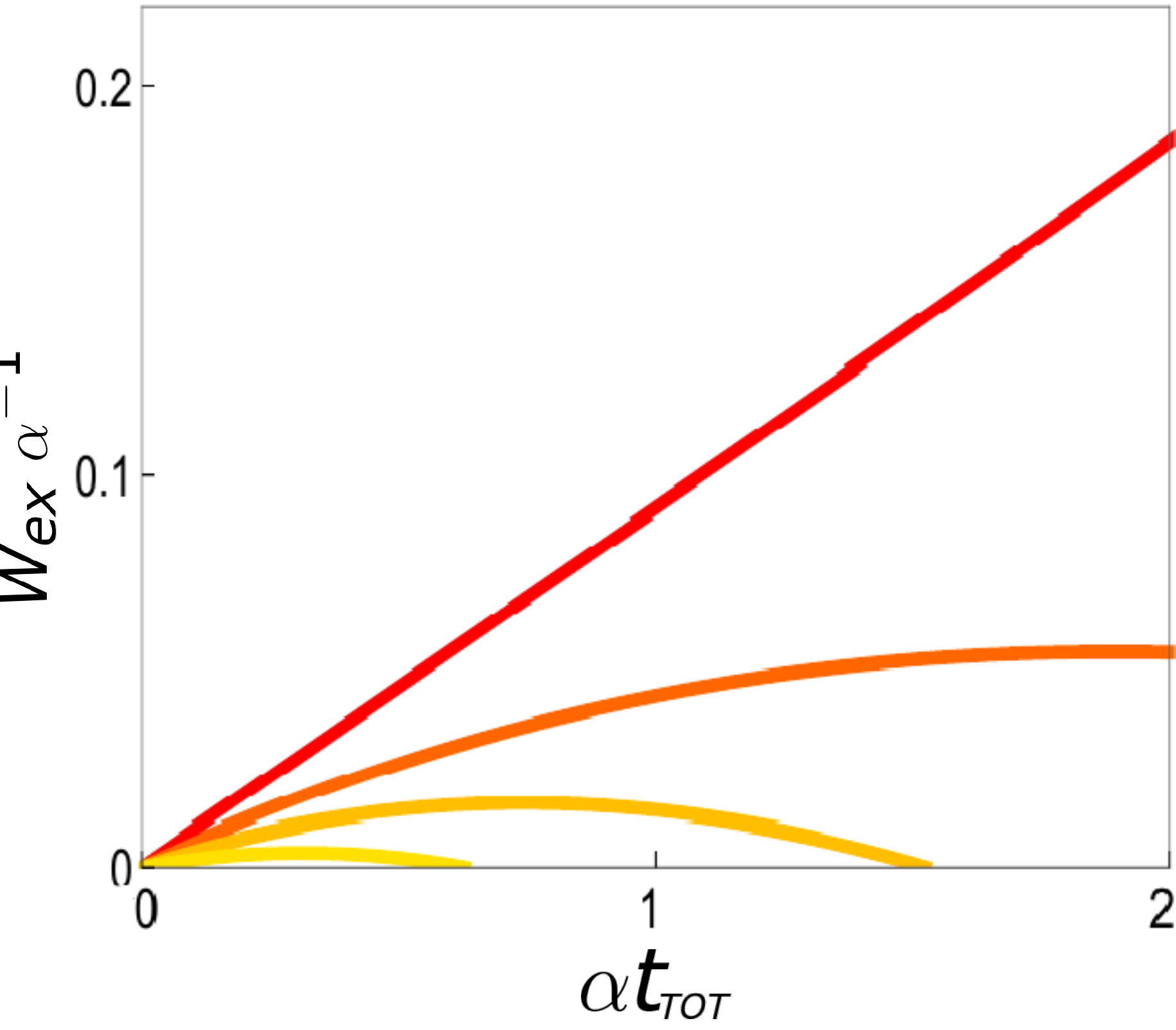}
\label{fig:W-t(beta)}} \subfigure[Power (${\cal P}$) vs. $\alpha
t_{tot}$ for various of $\theta$] {
\includegraphics[height=4cm,width=.42\columnwidth]{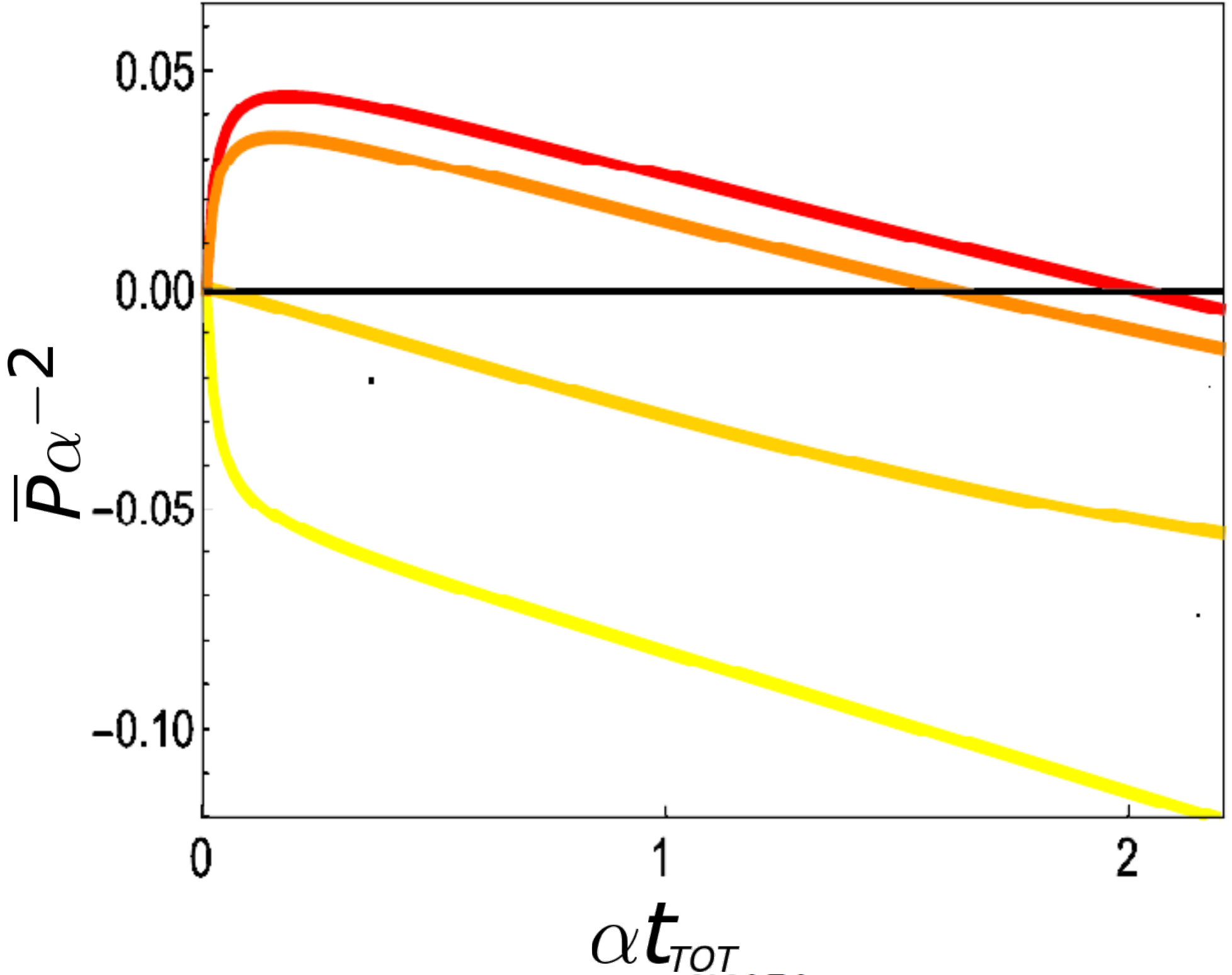}
\label{fig:P-t(theta)}} \subfigure[Power (${\cal P}$) vs. $\alpha
t_{tot}$ for various of $\beta_h/\beta_c$] {
\includegraphics[height=4cm,width=.42\columnwidth]{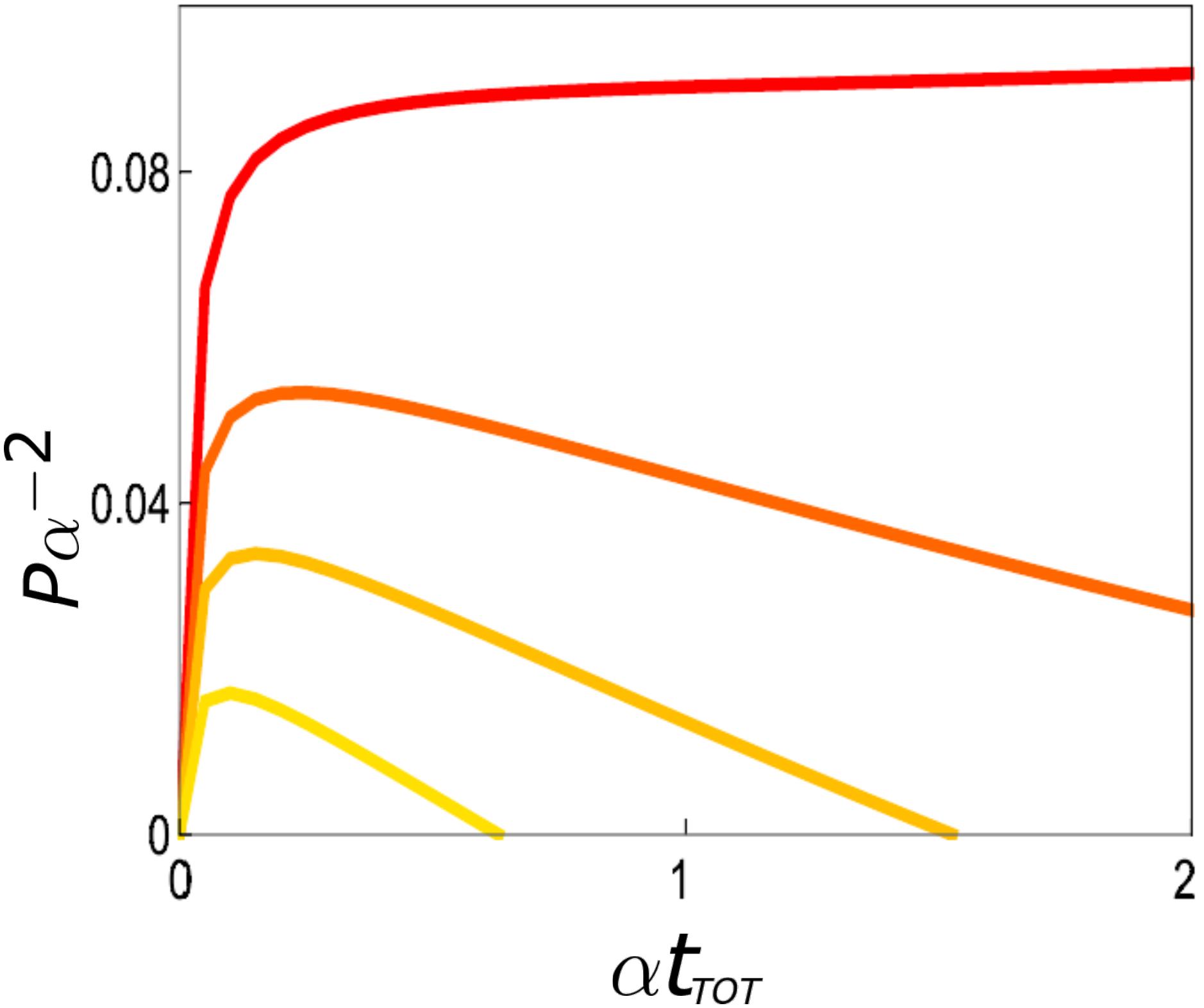}
\label{fig:P-t(beta)}} \subfigure[Efficiency ($\eta$) vs. $\alpha
t_{tot}$ for various $\theta$] {
\includegraphics[height=4cm,width=.42\columnwidth]{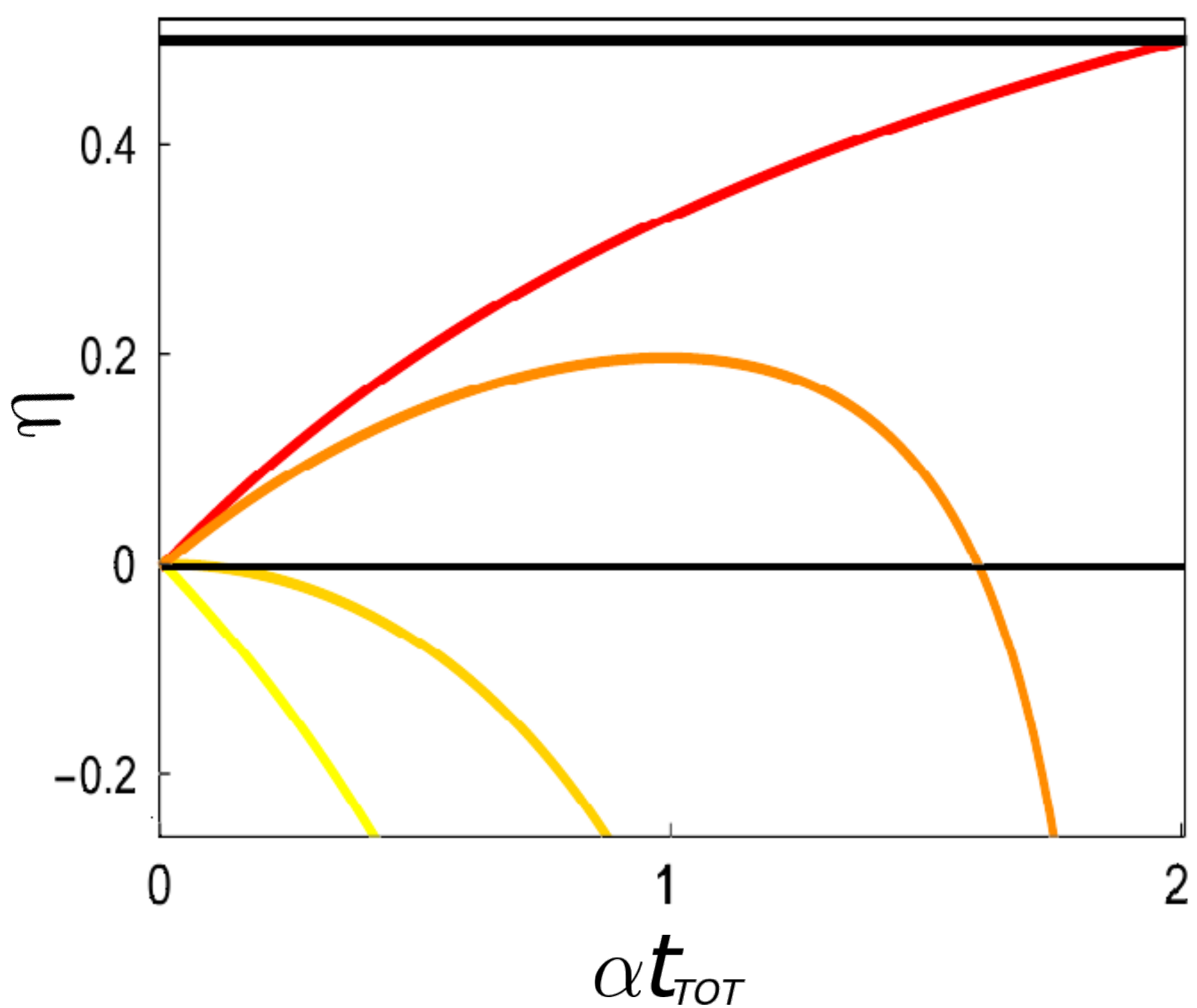}
\label{fig:eta-t(theta)}} \subfigure[Efficiency ($\eta$) vs.
$\alpha t_{tot}$ for various $\beta_h/\beta_c$] {
\includegraphics[height=4cm,width=.42\columnwidth]{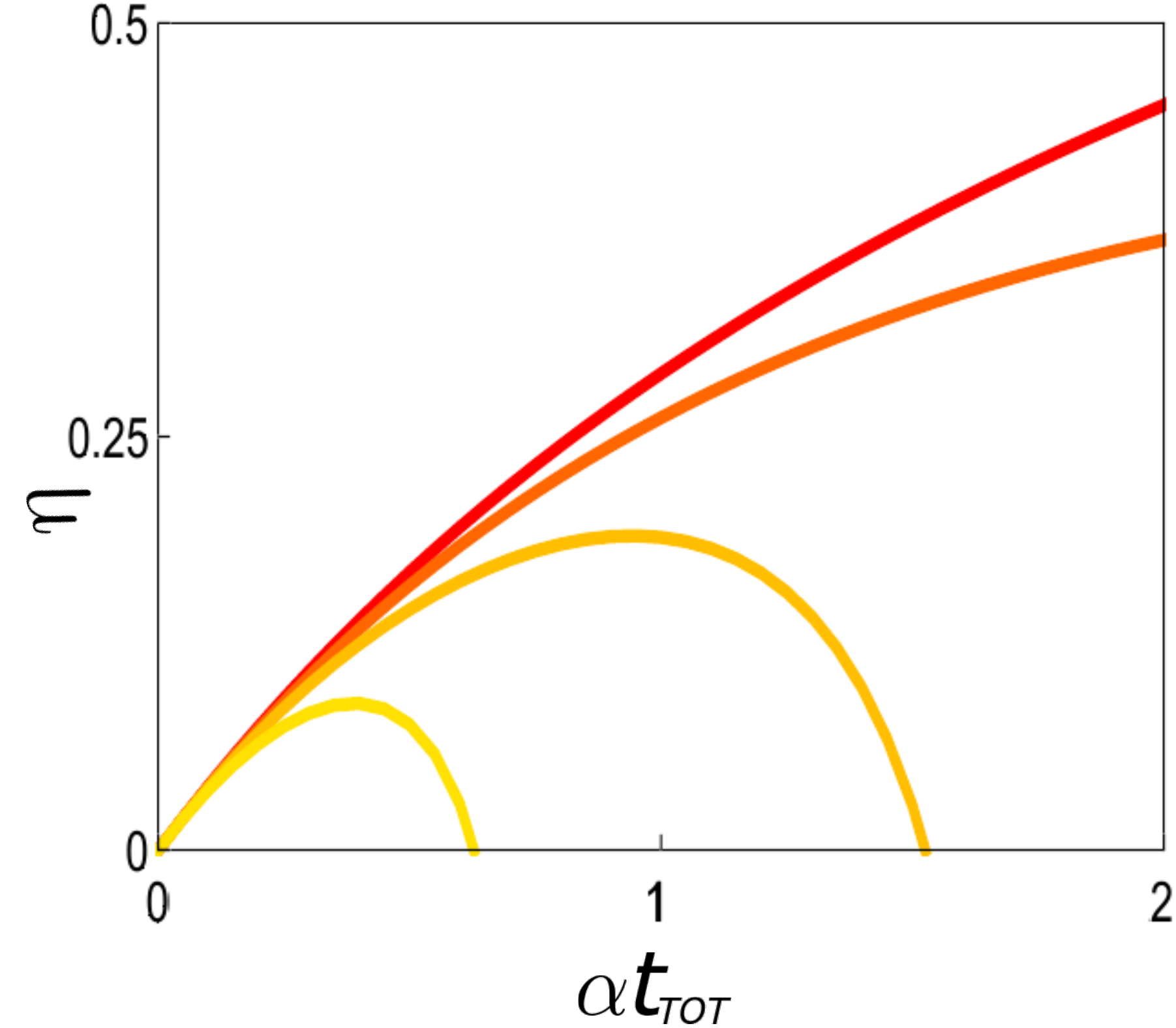}
\label{fig:eta-t(beta)}} \caption{Extractable work, power and
efficiency as functions of $\alpha t_{tot}$. In
\figurename\,\ref{fig:W-t(theta)}, \ref{fig:P-t(theta)} and
\ref{fig:eta-t(theta)} we fix the temperature ratio
$\beta_h/\beta_c=0.5$ and vary the misalignment $\theta$, starting
from $\theta=0$ for the highest (red) plot, then considering
$\theta=\pi/5, \pi/2$ and finally  $\theta= \pi$ for the lowest
(yellow) curve. On the other side, in \ref{fig:W-t(beta)},
\ref{fig:P-t(beta)} and \ref{fig:eta-t(beta)} we fix the
misalignment to $\theta=\pi/5$ and vary $\beta_h/\beta_c$, which
again going from the top-most red plot down to the yellow curve
takes the values $\beta_h/\beta_c=0.01, 0.31, 0.51, 0.71$.}
\label{fig:WPeta}
\end{figure}

\subsection{Extractable work, power and efficiency}

In what follows we will study the performance of the QOC in
various cases of equal rates for both of the adiabatic branches.
In \figurename\;\ref{fig:W-t(theta)}, \ref{fig:P-t(theta)} and
\ref{fig:eta-t(theta)} we plot $W_{ex}$, $\mathcal{P}$ and $\eta$
as function of the total time of the cycle $t_{tot}$ for different
values of the misalignment angle $\theta$ at a fixed value of the
ratio between the temperatures of the hot and cold reservoirs
$\beta_h/\beta_c=0.5$. It can be seen that the extractable work
becomes negative if the $t_{tot}$ exceeds a maximum time
$t_{M}(\theta)$, which is a function of the misalignment $\theta$.
This means that if the cycle lasts too long we are actually doing
work on the system. Moreover there exists a critical value of
$\theta$ such that the extractable work is negative for any value
of $t_{tot}$. Under these conditions, the cycle is not a heat
engine but rather a refrigerator, which uses external work to cool
the cold reservoir. An analogous behavior is shown by power and
efficiency.

%
%

In \figurename\, \ref{fig:W-t(beta)}, \ref{fig:P-t(beta)} and
\ref{fig:eta-t(beta)}, we show the dependence of $W_{ex}$,
$\mathcal{P}$ and $\eta$ on the total time of the cycle, for a
fixed misalignment angle $\theta=\pi/5$ and for different values
of the ratio $\beta_h/\beta_c$. We can see that as the ratio
increases the extractable work increases too. This is something
which is expected, nevertheless we can clearly see that the
finiteness in time of the cycle introduces again negative works
for $t_{tot}>t_{M}(\beta_h/\beta_c)$. This is again due to the
generation of inner friction, which comes along with the finite
time condition.

\begin{figure}
\centering
\includegraphics[height=4cm,width=.42\columnwidth]{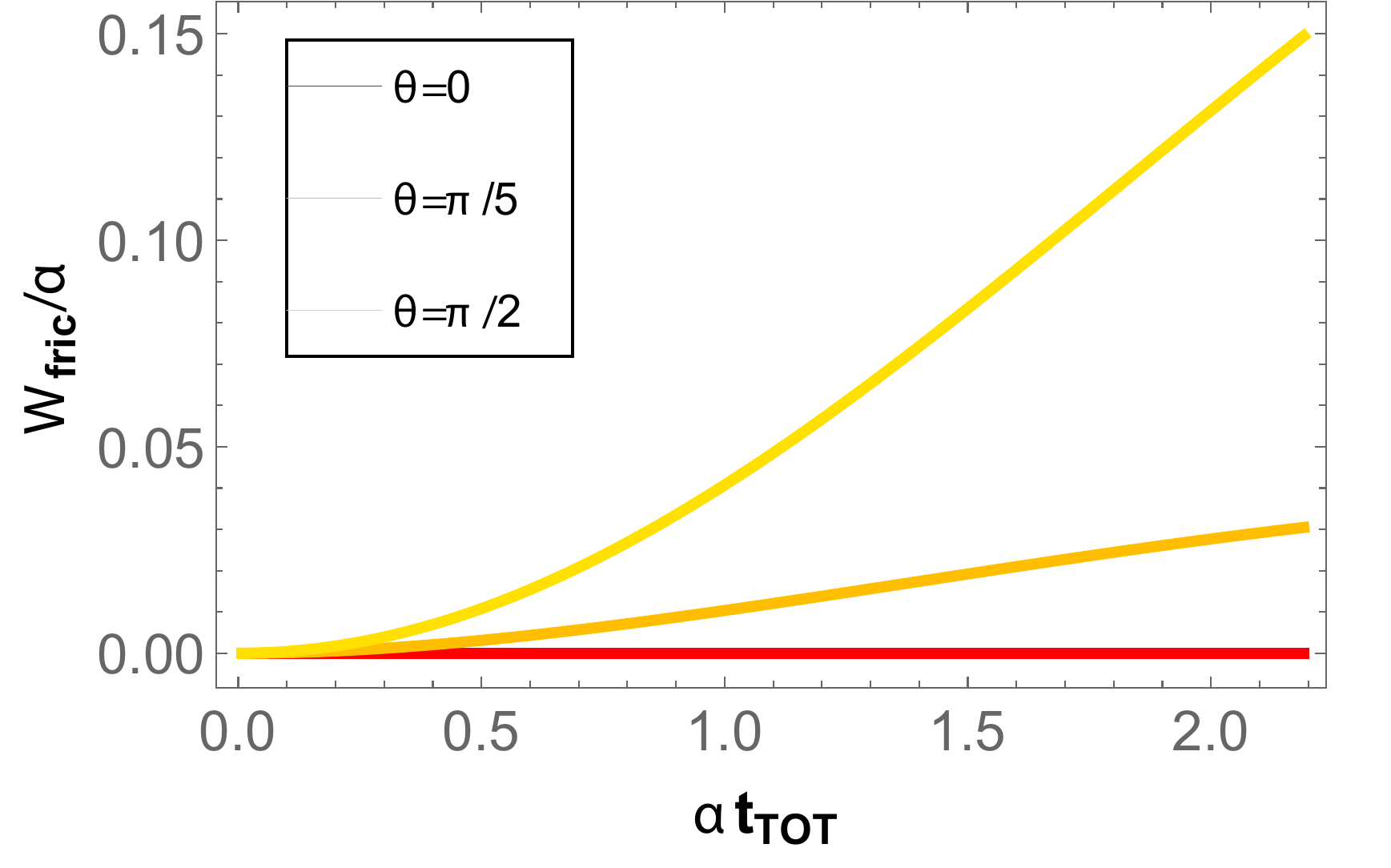}
\caption{Inner friction accumulated in the cycle as a function of
the total operation time for different misalignments $\theta$, at
$\beta_h/\beta_c=0.5$.} \label{fig:innfr}
\end{figure}
Inner friction is explicitly shown in \figurename\,
\ref{fig:innfr}, where the sum of the friction produced in the two
adiabatic strokes is shown as a function of the total cycle time
for various misalignment angles $\theta$. As discussed above, the
case $\theta=0$ is very special as no friction is generated,
whatever rate of variation is considered for the driving field
$\lambda(t)$. On the other hand, $W_{fric}$ increases with
increasing the tilting angle $\theta$ and decreases with
decreasing the driving rate $\alpha$. The behavior of $W_{fric}$
should be compared to that of $W_{ex}$ shown in \figurename \,
\ref{fig:WPeta}: the more friction is present, the less work can
be extracted from the engine.

\begin{figure}
\centering \subfigure[Average extractable work ($\overline
W_{ex}$) vs. $\alpha t_{tot}$ for various $\sigma$] {
\includegraphics[height=4cm,width=.42\columnwidth]{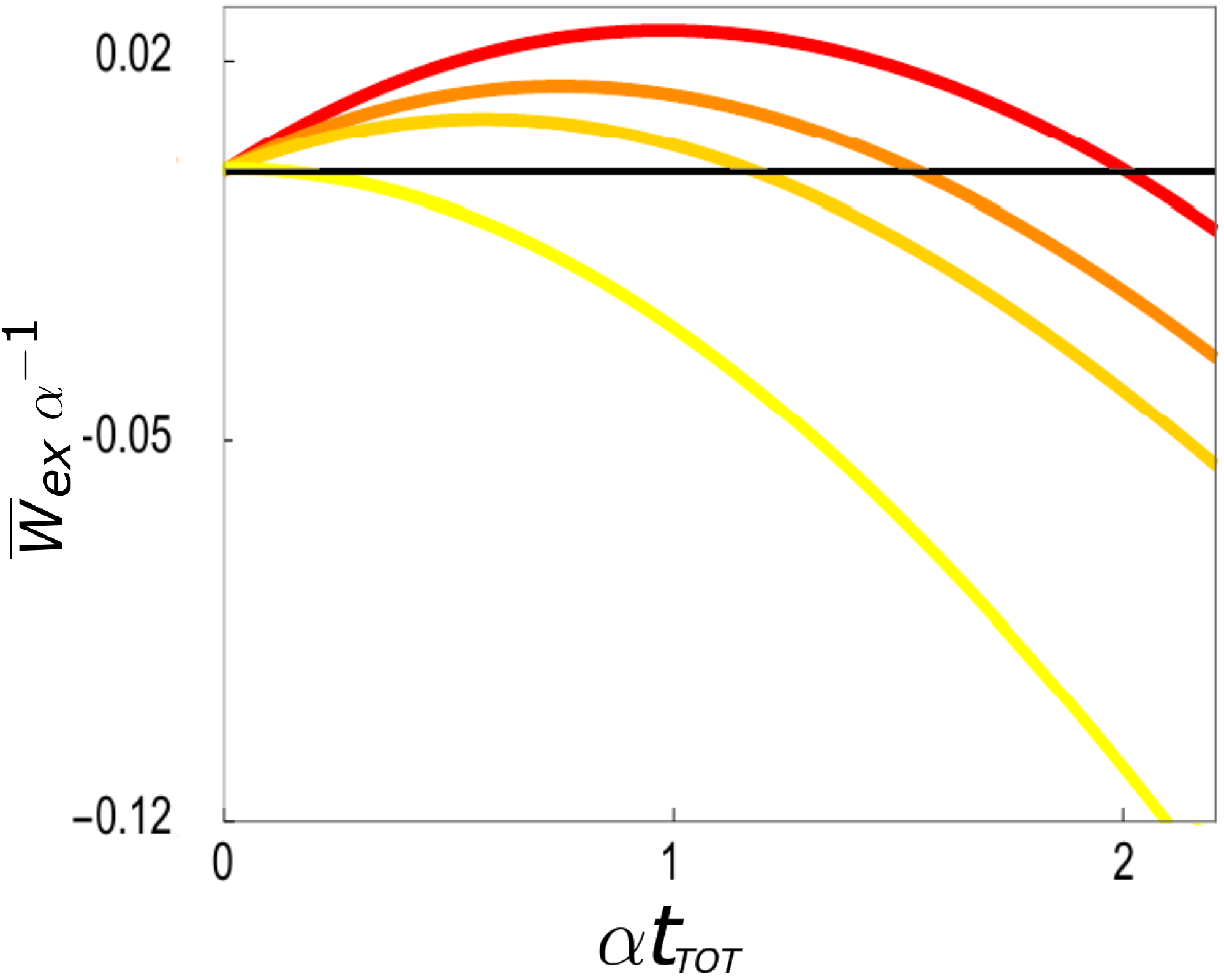}
\label{fig:W-t(sigma)}} \subfigure[Average extractable work
($\overline W_{ex}$) vs. $\alpha t_{tot}$ for various $\beta_h/\beta_c$]
{
\includegraphics[height=4cm,width=.42\columnwidth]{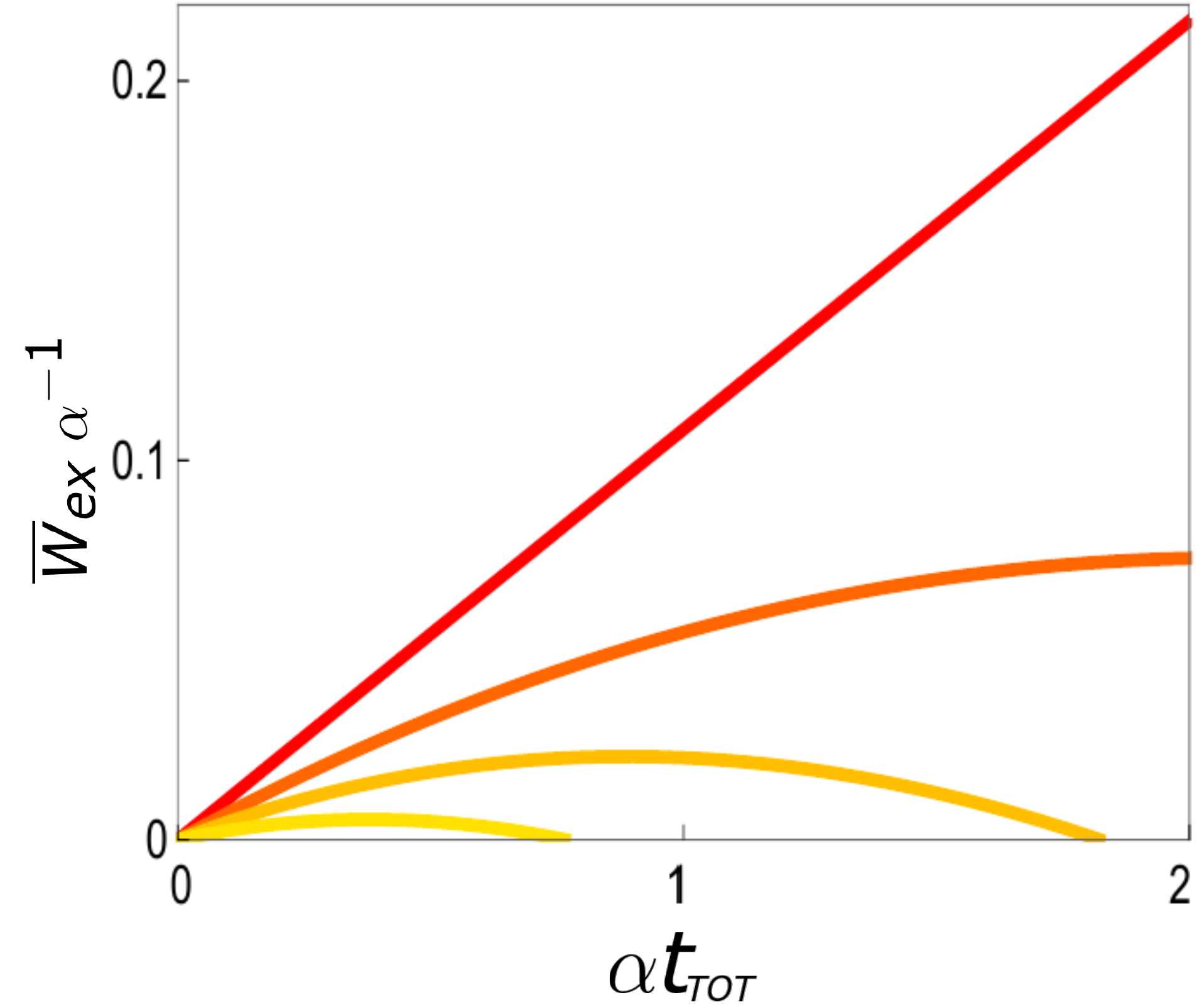}
\label{fig:W-taverage(beta)}} \subfigure[Averaged power
($\overline{\mathcal{P}}$) vs. $\alpha t_{tot}$ for various $\sigma$] {
\includegraphics[height=4cm,width=.42\columnwidth]{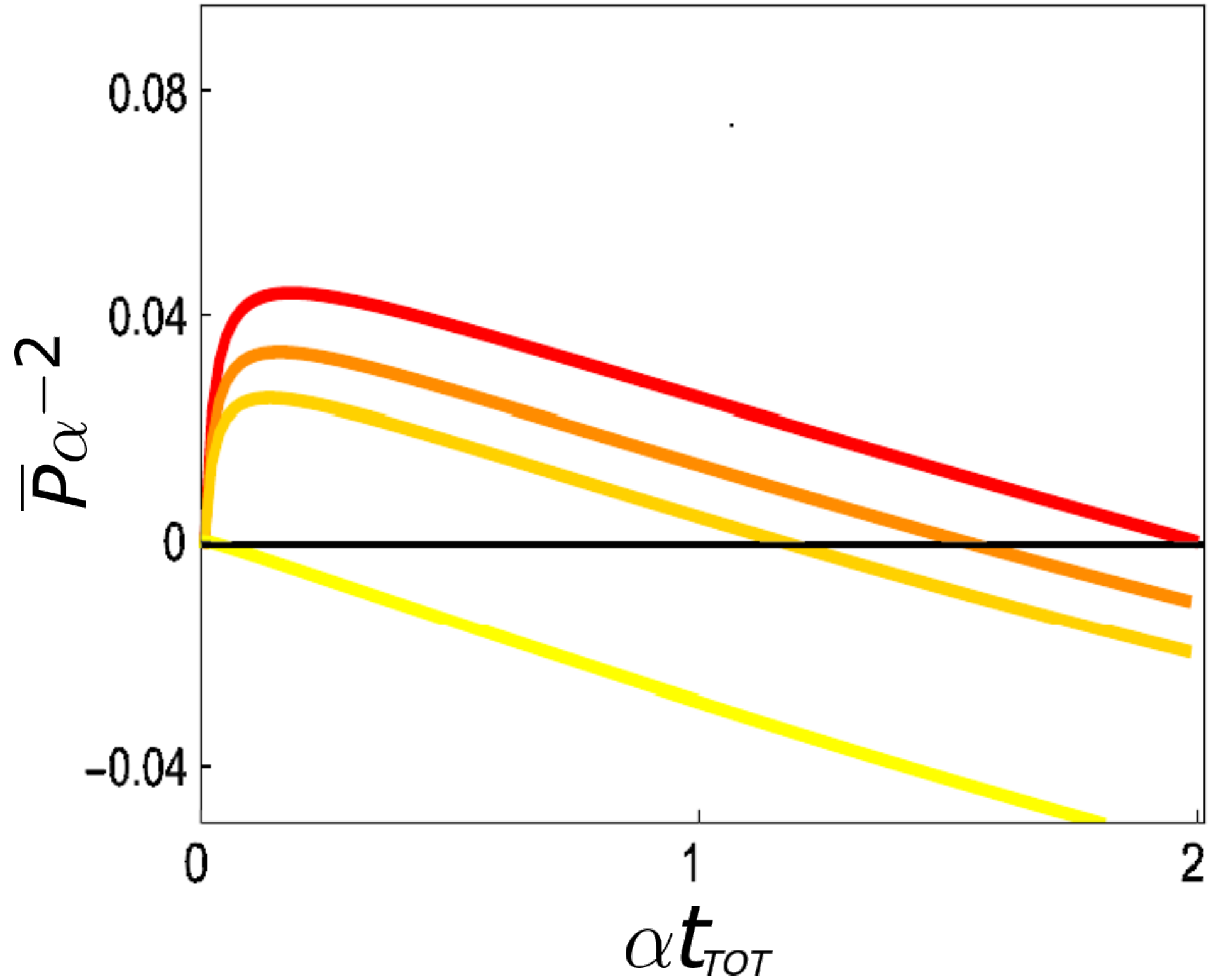}
\label{fig:P-t(sigma)}} \subfigure[Averaged power
($\overline{\mathcal{P}}$) vs. $\alpha t_{tot}$ for various
$\beta_h/\beta_c$] {
\includegraphics[height=4cm,width=.42\columnwidth]{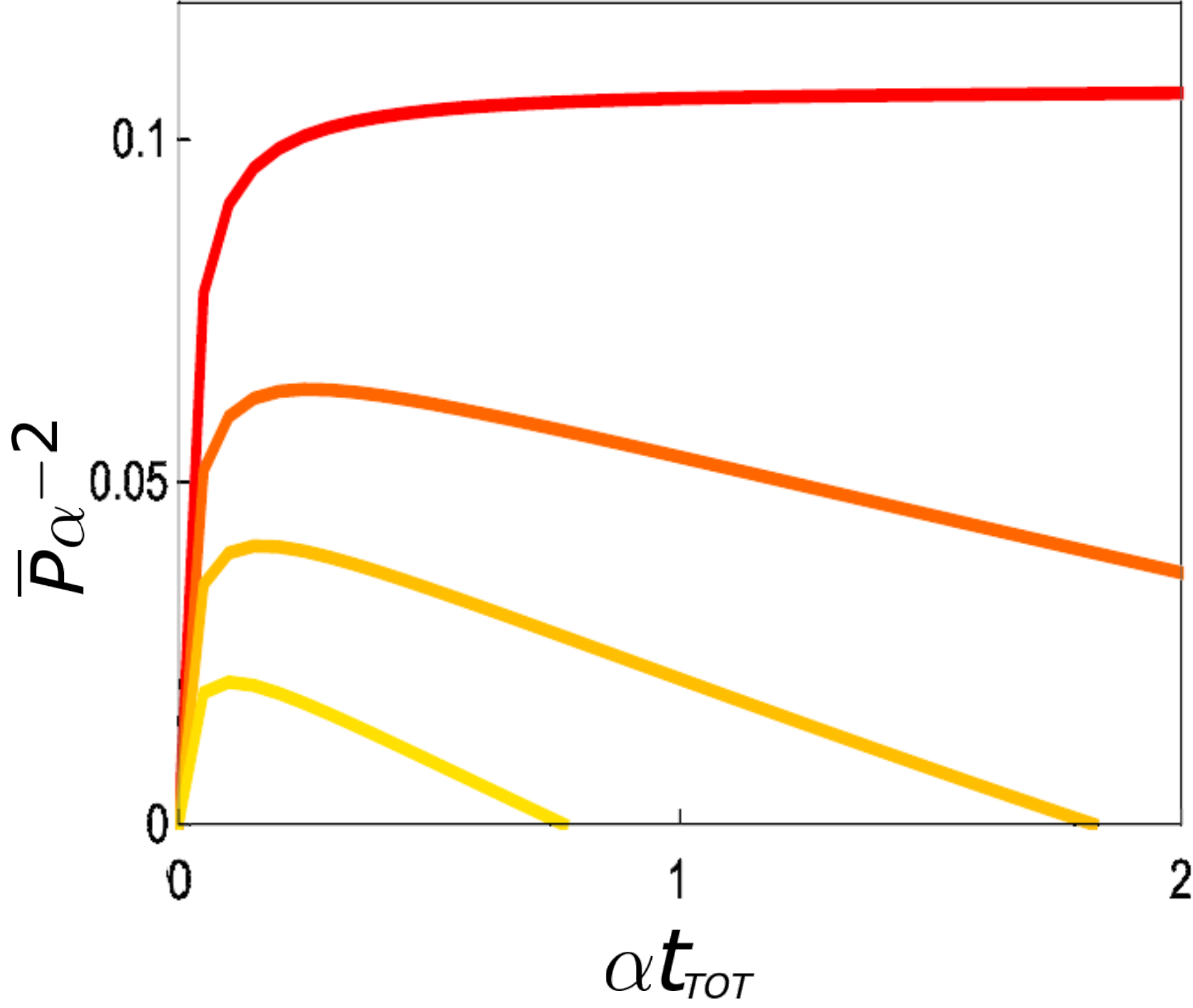}
\label{fig:P-taverage(beta)}} \subfigure[Averaged efficiency
($\overline{\eta}$) vs. $\alpha t_{tot}$ for various $\sigma$]
{\includegraphics[height=4cm,width=.42\columnwidth]{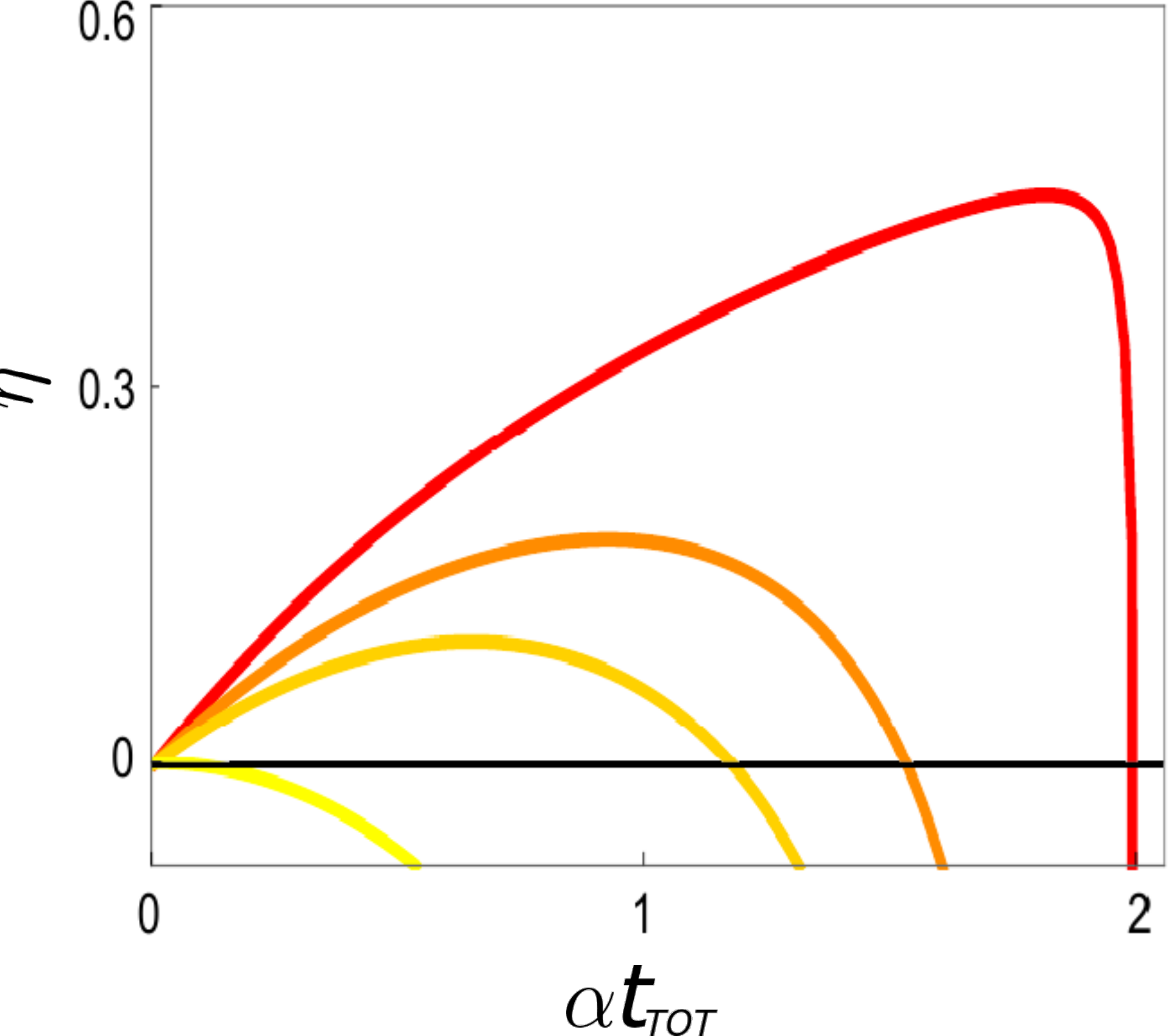}
\label{fig:eta-t(sigma)} } \subfigure[Averaged efficiency
($\overline{\eta}$) vs. $\alpha t_{tot}$ for various $\beta_h/\beta_c$]
{\includegraphics[height=4cm,width=.42\columnwidth]{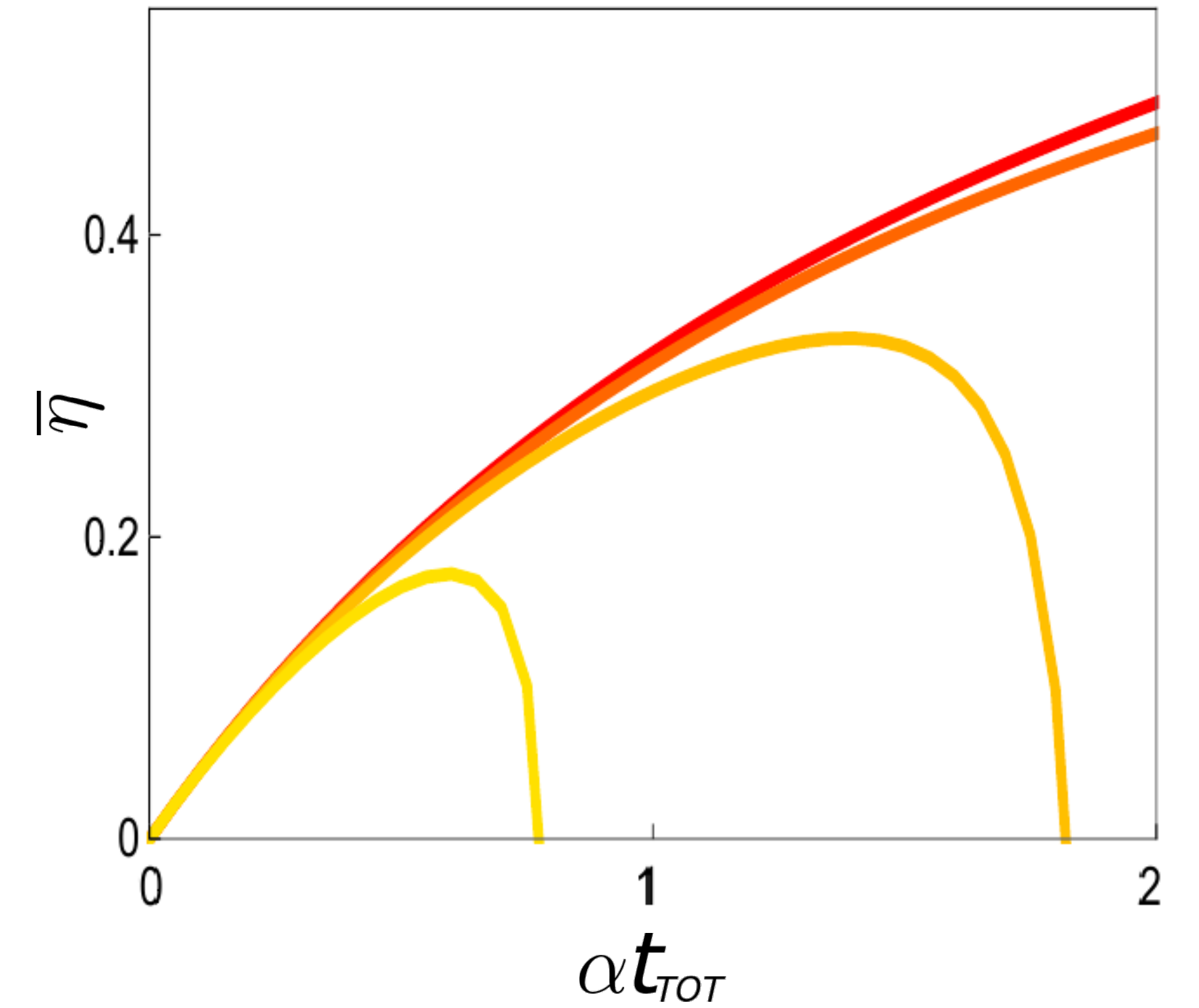}
\label{fig:eta-taverage(beta)}} \caption{Averaged extractable
work, power and efficiency as functions of $\alpha t_{tot}$.
In \figurename\, \ref{fig:W-t(sigma)}, \ref{fig:P-t(sigma)} and
\ref{fig:eta-t(sigma)},  we consider a fixed
value for the temperature ( $\beta_h/\beta_c=0.5$ ) and vary
$\sigma$. Different colors refer to different gaussian
widths: the red plots correspond to $\sigma^2=0.01$, the orange
ones to $\sigma^2=0.5$, light orange ones to $\sigma^2=1$ and
finally the (lowest) yellow curves refer to a flat distribution.
In
\ref{fig:W-taverage(beta)}, \ref{fig:P-taverage(beta)} and
\ref{fig:eta-taverage(beta)}  we take $\sigma^2=0.1$ and vary
$\beta_h/\beta_c$, which, going from the red plots to the yellow
ones, takes the values $\beta_h/\beta_c=0.01, 0.31, 0.51, 0.71$.
}
\label{fig:WPetaaverage}
\end{figure}

So far we have considered QOCs with given values of the
misalignment angle $\theta$. We now consider the effect of disorder and
assume that $\theta$ is a
Gaussian random variable with mean value $\overline\theta=0$ and
variance $\sigma^2$. In \figurename\, \ref{fig:W-t(sigma)},
\ref{fig:P-t(sigma)} and \ref{fig:eta-t(sigma)} we show the
behavior of extractable work, power and efficiency for different
values of the variance and given temperature ratio
$(\beta_h/\beta_c=0.5)$. We can see that at a given total time
$t_{tot}$, the best performance is always obtained with sharper
distributions (smaller $\sigma$). Thus, if the disorder of the
system grows, the capability of the latter of providing work and
of doing it in a more efficient way decreases. Again we mention
the fact that there exist a maximum total time $t_{M}$ above which
the QOC is not an heat engine anymore. We also notice that even a small
disorder has quite dramatic effect in reducing the efficiency for long enough times
(upper curve in Fig.\ref{fig:eta-t(sigma)}).

In \figurename\, \ref{fig:W-taverage(beta)},
\ref{fig:P-taverage(beta)} and \ref{fig:eta-taverage(beta)} we
plotted the behavior of the same quantities for different values
of the ratio $\beta_h/\beta_c$ at a given variance $\sigma^2=0.1$.
Again, all of the quantities increase as the difference in
temperatures increases.

\begin{figure}
\centering \subfigure{
\includegraphics[height=4cm,width=.42\columnwidth]{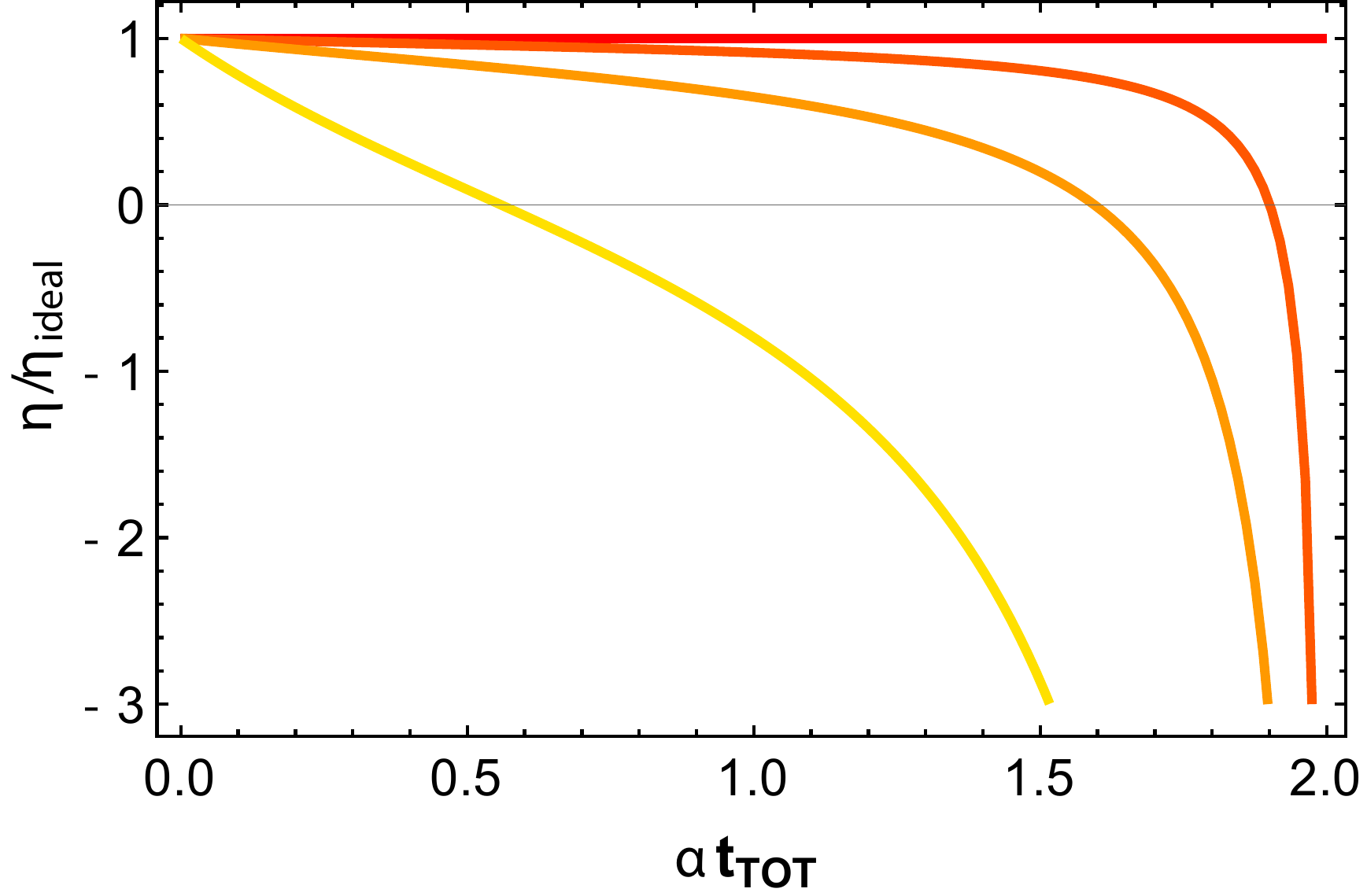}
\label{fig:nuova1-theta}} \subfigure{
\includegraphics[height=4cm,width=.42\columnwidth]{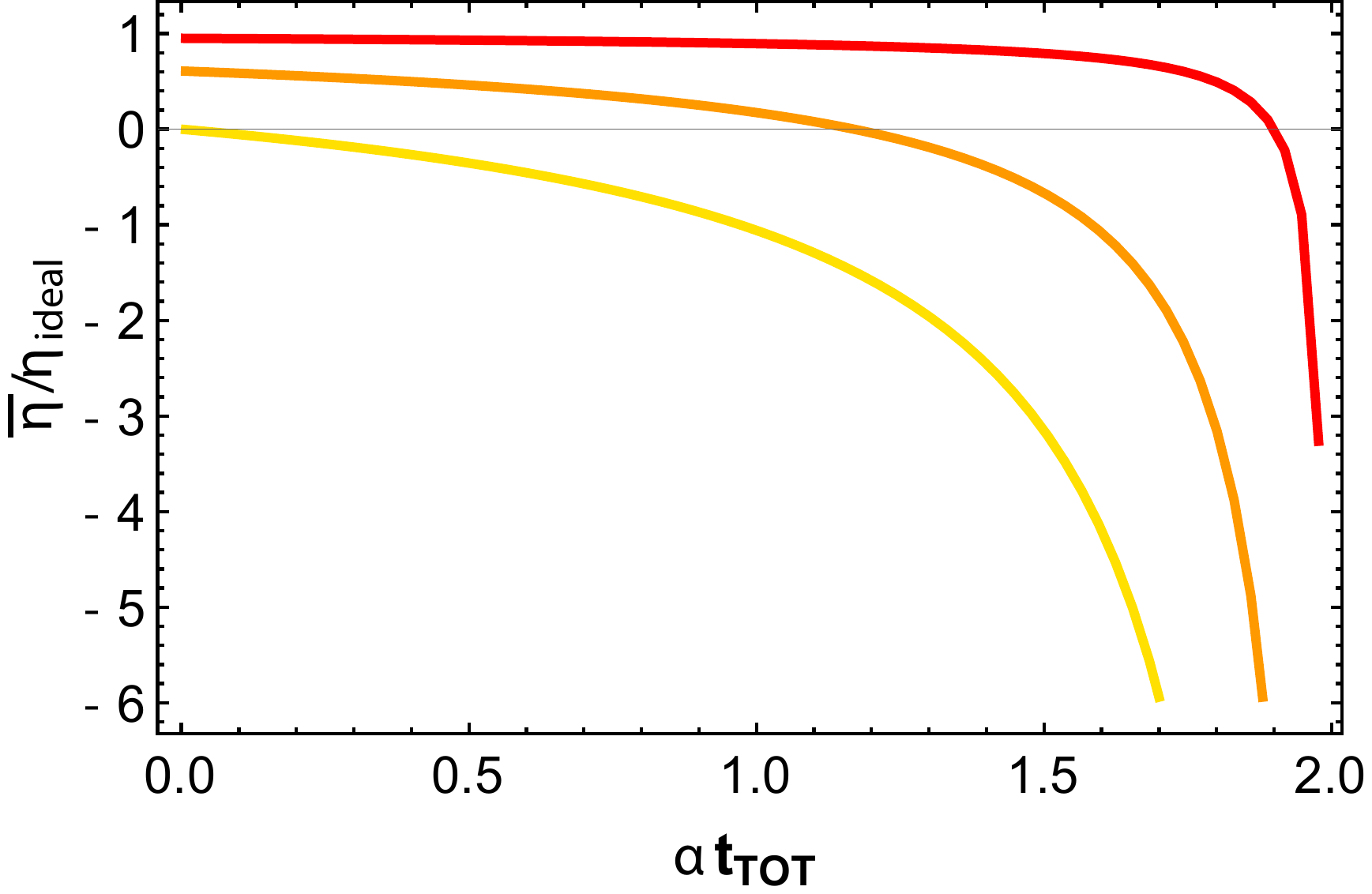}
\label{fig:nuova2-average}} \caption{Left: renormalized efficiency
$\eta/\eta_{ideal}$ as a function of the total operation time for
different misalignments $\theta = 0, \pi/10, \pi/5, 2 \pi/5$, at
$\beta_h/\beta_c=0.5$. Right: averaged efficiency $\bar{\eta}$
normalized with respect to the ideal efficiency obtained at
$\theta=0$. The average is taken over gaussian distributions with
variances, $\sigma^2 = 0.1$, $\sigma^2= 1$, and over a flat
distribution, respectively. For all of the plots, we fixed the
temperature ratio $\beta_h/\beta_c=0.5$. } \label{fig:nuova}
\end{figure}

We plotted all of the quantities as a function of the total cycle
time $t_{tot}$ because, from an operational point of view, this is
the quantity one can control once the working substance is
prepared and the stage is set for the thermal machine to operate.
However, it has to be mentioned that some care should be payed
when comparing the values of the efficiency at different operating
times. Indeed, since $\omega$ varies with time, each different
$t_{tot}$ gives rise to a different value of the final frequency
(called $\omega_2$, above) at which the isochoric $2 \rightarrow
3$ takes place, see \figurename\, (\ref{fig:phasespace}). The
ideal cycle with infinitely slow adiabatic branches, corresponding
for us also to the absence of misalignment ($\theta=0$) and shown
for comparison in each plot in \figurename\, (\ref{fig:WPeta}) and
(\ref{fig:innfr}), has efficiency $\eta_{ideal} = 1-
\frac{\omega_1}{\omega_2}$. The dependence of $\eta_{ideal}$ on
$\omega_2$ implies that the efficiency $\eta$  should be compared
with a different $\eta_{ideal}$ at each different $t_{tot}$. To
avoid any confusion in this respect, and to better display the
role of finite-time induced friction in the machine performance,
we show this comparison in \figurename\, (\ref{fig:nuova}), where
the ratio $\eta/\eta_{ideal}$ is displayed as a function of the
operating time. Once the efficiency is renormalized in this way,
its residual dependence on $t_{tot}$ can be fully ascribed to the
presence of inner friction.

\subsection{Efficiency at maximum power}
\begin{figure}
\centering \subfigure[$\mathcal{P}$ vs. $\eta$ for various
$\sigma$] {
\includegraphics[height=5cm,width=.42\columnwidth]{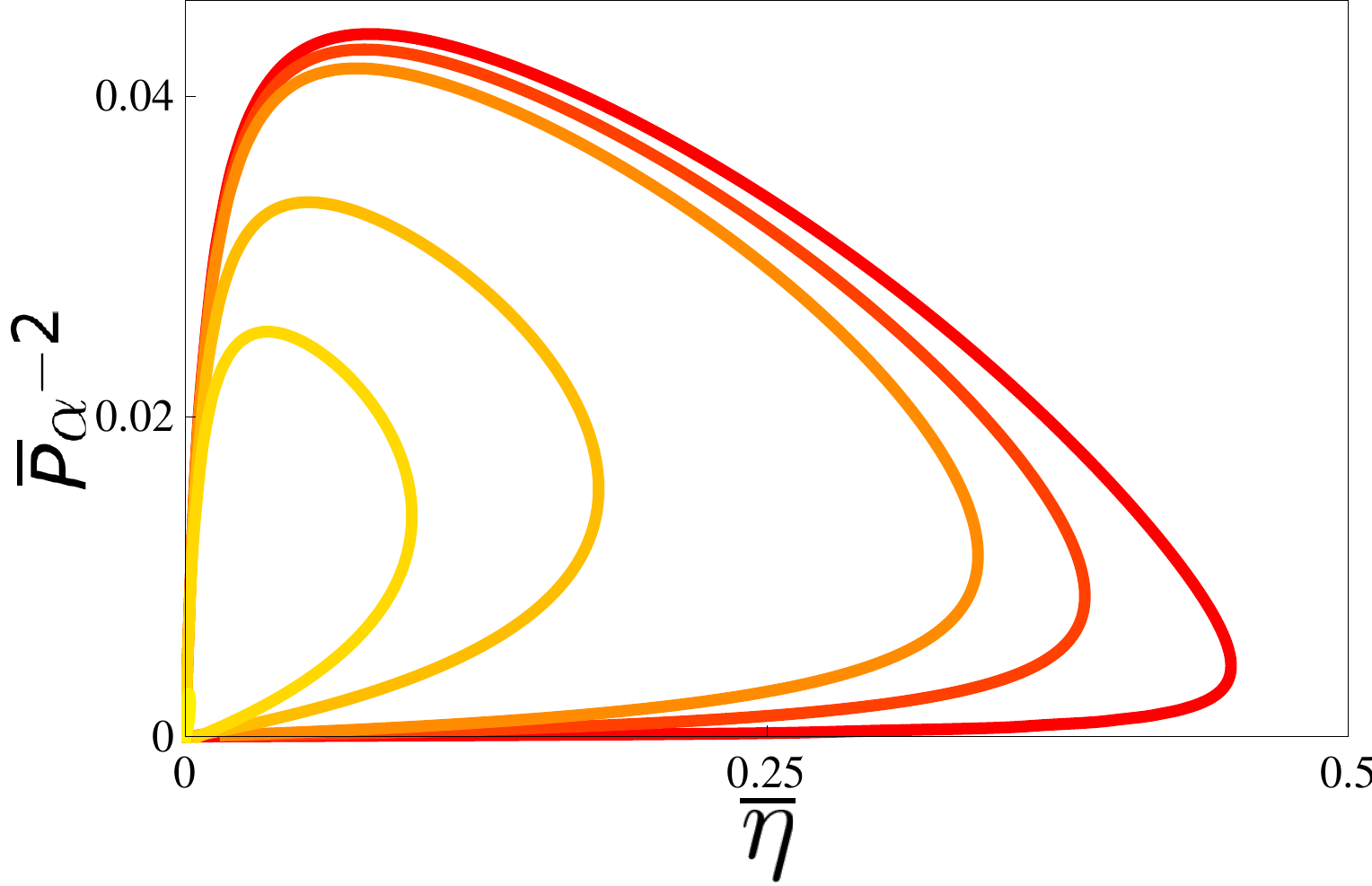}
\label{fig:Petasigma}} \subfigure[$\mathcal{P}$ vs.$\eta$ for
various $\beta_h/\beta_c$] {
\includegraphics[height=5cm,width=.42\columnwidth]{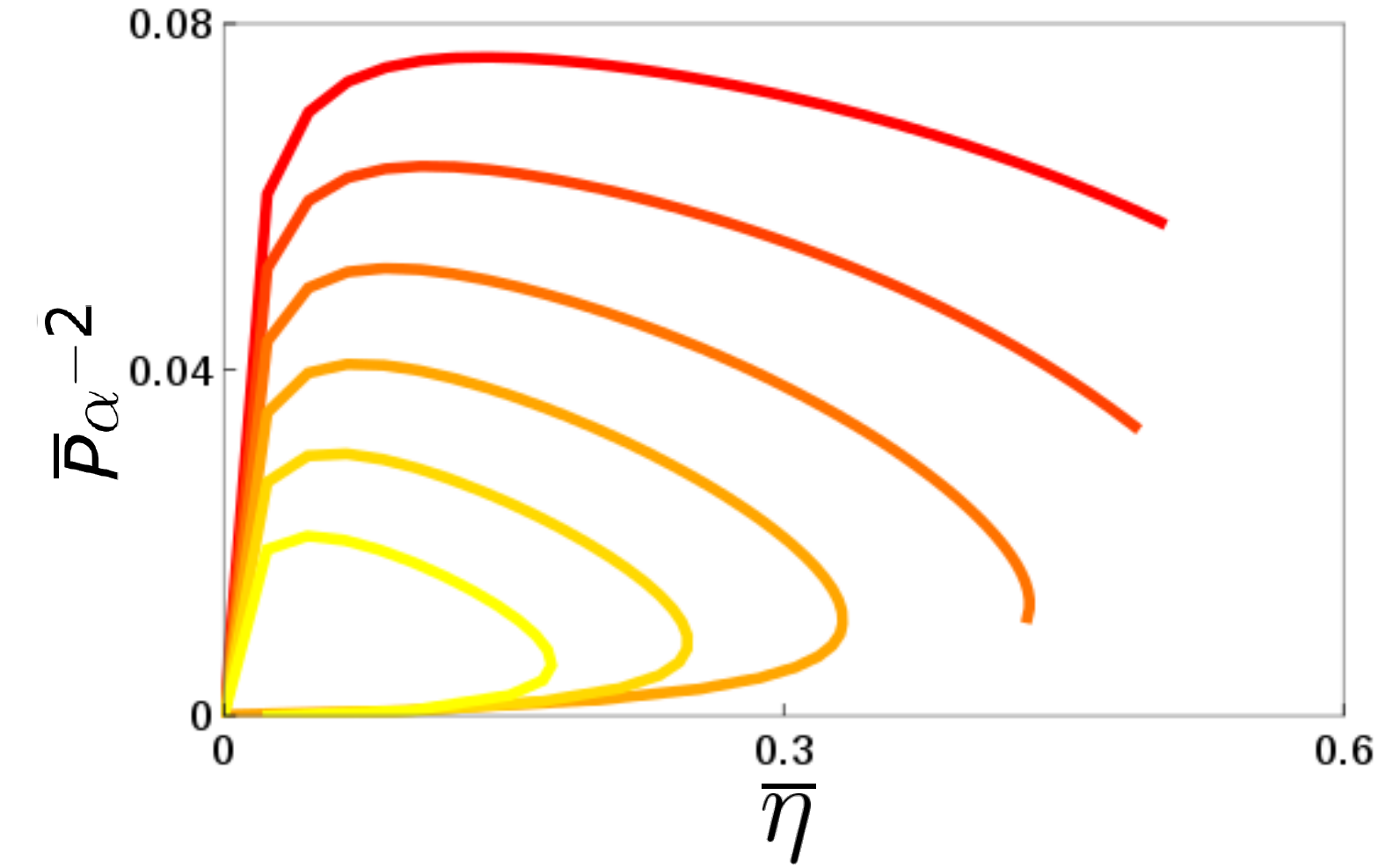}
\label{fig:Petabeta}} \caption{Relation between averaged power and
efficiency, $\overline{\mathcal{P}}(t)$ and $\overline{\eta}(t)$ at the same time parameter $\alpha t_{tot}$. In
\ref{fig:Petasigma} we choose $\beta_h/\beta_c=0.5$ and vary
$\sigma$, which, from the outer to the inner curve takes the
values $\sigma^2=0.01, 0.05, 0.1, 0.5, 1, 10$. In
\ref{fig:Petabeta} we fix $\sigma^2=0.1$ and starting from the
outer to the inner curve we consider the increasing temperature
ratios $\beta_h/\beta_c=0.02,0.03,0.04,0.05,0.06,0.07$.}
\label{fig:Peta}
\end{figure}

Let us now consider the relation between $\eta$ and $\mathcal{P}$
(see \figurename \ref{fig:Peta}) and then extract the value of the
efficiency at maximum power $\eta(\mathcal{P}_{MAX})$. Two sets of
these data are reported in  Tab. \ref{tab:etaPmaxsigma} and
\ref{tab:etaPmaxbeta}. They refer to averaged power and efficiency
considering in the first case various temperature ratios at a
fixed width, and the other way round for the second case. The
effects of disorder and temperature difference continue to stand:
our analysis provides larger values of (averaged) power and
(averaged) efficiency at maximum power for smaller and smaller
$\sigma$ and for larger and larger $\beta_h/\beta_c$, with the
following one-sentence summary: We obtain considerably larger
values of $\eta (\mathcal{P}_{MAX})$ when the temperatures ratio
is large and for very picked misalignment $G$ distributions, that
is, when the inner friction is smaller.

\begin{table}
\label{tab:etaPmaxsigma}
\centering
\renewcommand\arraystretch{1.3}
\begin{tabular}{>{$}c<{$}>{$}c<{$}>{$}c<{$}>{$}c<{$}} \br
\sigma^2 & \alpha t_{tot}^{MAX} &
\overline{\mathcal{P}}_{MAX}/\alpha^{2} &
\overline{\eta}(\overline{\mathcal{P}}_{MAX}) \\ \mr
0.01  & 0.0882 & 0.0439 & 0.0775 \\
\hline
0.05  &  0.0882 & 0.0429 & 0.0758\\
\hline
0.1   &  0.0882 & 0.0418 & 0.0737\\
\hline
0.5   &  0.0771 & 0.0334 & 0.0519\\
\hline
1               &  0.0340 & 0.0253 & 0.0340\\
\hline
10    &  0.0.220 & 0.0027 & 0.0220\\
\end{tabular}
\caption{Efficiency at maximum power at $\beta_h/\beta_c=0.5$ and
for different values of the Gaussian bell's width $\sigma$. The
optimal total cycle time, $\alpha t_{tot}^{MAX}$, is the one for
which $\overline{\mathcal{P}}$ attains its maximum.}
\end{table}

\begin{table}
\centering
\renewcommand\arraystretch{1.3}
\label{tab:etaPmaxbeta}
\begin{tabular}{>{$}c<{$}>{$}c<{$}>{$}c<{$}>{$}c<{$}} \br
\beta_h & \alpha t_{tot}^{MAX} &
\overline{\mathcal{P}}_{MAX}/\alpha^{2} &
\overline{\eta}(\overline{\mathcal{P}}_{MAX}) \\ \mr
2.1  & 0.175 & 0.0761 & 0.1420 \\
\hline
3.1  &  0.125 & 0.0635 & 0.1056\\
\hline
4.1   &  0.1 & 0.0517 & 0.0862\\
\hline
5.1   &  0.075 & 0.0406 & 0.0660\\
\hline
7.1   &  0.05 & 0.0208 & 0.0447\\
\hline
9.1  &  0.025 & 0.0045 & 0.0224\\
\br
\end{tabular}
\caption{Efficiency at maximum power for $\sigma^2=0.1$ for
different temperature ratios $\beta_h/\beta_c$. The maximum power
$\overline{\mathcal{P}}_{MAX}=\overline{\mathcal{P}}(t_{tot}^{MAX})$
is obtained for the times $\alpha t_{tot}^{MAX}$ reported in the
second column.}
\end{table}

\section{Experimental implementation with an optical set-up}
\label{sec:exp}

 In this Section we propose one possible experimental set-up
 by means of which it is possible to realize the Otto cycle discussed so far
 and test our findings.
 The Otto cycle is made up of two different types of branches,
 namely adiabatic and isochoric transformation.
 Thus the proposed set-up has to be able to implement both of them.

 The physical system we have in mind is an optical one and in particular
 we propose to encode the qubit into the polarization degree of freedom of a single
 photon.
 In the following we address the implementation of the two types of branches
 separately, stressing the key points for both of them.

\subsection{Implementation of the adiabatic transformation}

  The adiabatic branch in the Otto cycle is achieved by means
  of a time dependent effective magnetic field and its time evolution is given
  by the unitary operator:

  \begin{equation}
  \label{eq:unitary}
   \hat{U}(t)=T e^{-\imath \int_{0}^{t}d\tau\;\; \vec{B}(\tau) \cdot \vec{\sigma}}
  \end{equation}

  For a fixed $t=t^{*}$ the above operator can be written as a rotation in the Hilbert space of the qubit using the Euler decomposition as:
 \begin{equation}
  \label{eq:euler}
   \hat{U}(t^{*})=e^{-\imath \frac{\psi^{*}}{2} \sigma_z}e^{-\imath \frac{\theta^{*}}{2} \sigma_x}e^{-\imath \frac{\phi^{*}}{2} \sigma_z}.
  \end{equation}
 This expression is helpful for our purposes because the single rotations
 appearing in it can be easily implemented in an optical setup as rotation
 of the polarization degrees of freedom of a single photon.

 Therefore by encoding the qubit into the polarization degree of freedom of a photon and
 in particular by choosing the basis $\{\left|H\right\rangle,\left|V\right\rangle\}$
 of horizontal and vertical polarization, we can perform the wanted rotations by means of
 properly chosen phase retarders.

 \subsection{Implementation of the isochoric transformation}

 The isochoric transformation requires more care.
 By definition it amounts to attaching the system to a thermal bath,
 which makes the system thermalize into a Gibbs-like state.
 The latter is characterized by a density matrix $\rho$ with no coherences between
 different eigenstates of the Hamiltonian of the system whereas the diagonal
 ones are given by the Boltzmann factors $e^{-\beta \epsilon_n}/(e^{-\beta \epsilon_0}+e^{-\beta \epsilon_1})$,
 where $\beta=1/T_c$ or $\beta=1/T_h$ is the inverse temperature we want the state to thermalize at
 and $\epsilon_n$ are the eigenenergies of the final Hamiltonian of the
 adiabatic transformation preceding the isochoric one we are addressing.
 In the case of a qubit the thermalization process leads to
 a final state which is diagonal in a given basis (determined by the
 form of the bath-spin coupling) and whose populations are related to the temperature of the
 thermal bath by the relation
 \begin{equation}
  \label{eq:temperature}
  \beta=T^{-1}=-\frac{1}{\epsilon_1-\epsilon_0}\log{\frac{1-p_0^{(f)}}{p_0^{(f)}}}\;\;\;\; \frac{1}{2}\le p_0^{(f)}< 1,
 \end{equation}
 where, $p_0^{(f)}$ is the population of the
 lowest $(n=0)$ state of the qubit after thermalization has occurred.

 In order to implement an isochoric transformation we propose to exploit the experimental set-up
 used in Ref.~ \cite{photonic}. The idea is to exploit the spatial degrees of freedom
 of the photon as an effective bath for its polarization.
 The coupling between the two is achieved by exploiting the birefringent property of
 a quartz plate.
 The effect of the latter on a photon passing through it is to phase-shift the horizontal and vertical
 component of the polarization by an amount proportional to the number of photons per
 mode.
 Once the spatial part of the photon is traced out, the dynamics of the polarization turns out to
 be driven by the following dynamical map between an initial state $\rho_i$ to a
 final state $\rho_f$, which describes decoherence:
 \begin{equation}
 \label{eq:map}
   \rho_i\rightarrow\rho_f=\frac{1}{2}\left((1+z)\,\rho_i+(1-z)\,\hat{\sigma}_{z}\,\rho_i\,\hat{\sigma}_{z}\right),
  \end{equation}
 where the parameter $z$ can be tuned from $z=1$ (identity map) to $z=-1$ (complete decoherence, namely
 the final density matrix has vanishing off-diagonal terms).
 Because of our assumption of complete thermalization we will always assume $z=-1$.

 We can thus exploit this mechanism in order to engineer thermalization in the following way.
 Let us assume that the inverse temperature of the bath we want to mimic is $\beta$.
 Through Eq.~(\ref{eq:temperature}) we can determine
 the population of the lowest energy level after the system completely
 thermalized.
 Let us write the initial state (which in turn correspond to the final state of the adiabatic
 transformation preceding the isochoric one) as:
 \begin{equation}
  \label{eq:instate}
  \rho_i=\frac{1}{2}\mathbf{1}+\left(\frac{1}{2}-p_{0}^{(i)}\right)\sigma_z+b_x\; \sigma_x+b_y\;\sigma_y
 \end{equation}
 Since the decoherence mapping in Eq.~(\ref{eq:map}) has the effect of making the off diagonal elements
 vanish we first need to perform a rotation on the initial state $\rho_i$ to turn it into a state
 of the form:
 \begin{equation}
  \label{eq:mfinal}
  \rho_f'=\frac{1}{2}\mathbf{1}+\left(\frac{1}{2}-p_{0}^{(f)}\right)\sigma_z+b_x'\; \sigma_x+b_y'\;\sigma_y
 \end{equation}
 where $p_0^{(f)}$ is calculated through relation in Eq.~(\ref{eq:temperature}).
 The application Eq.~(\ref{eq:map}) has now the effect of making $b_x'=b_y'=0$,
 thus leaving us with the desired state:
 \begin{equation}
  \label{eq:final}
  \rho_f=\frac{1}{2}\mathbf{1}+\left(\frac{1}{2}-p_{0}^{(f)}\right)\sigma_z
 \end{equation}
 It is easy to see that in order to get from the state in Eq.~(\ref{eq:instate})
 to the one in Eq.~(\ref{eq:mfinal}) we can apply a specific rotation, which has to be chosen
 by taking into account both states.
 For instance in the case $b_y=0$ and $p_0^{(f)}<p_0^{(i)}\le1/2$,
 that is, if
 we are `heating' our system,  the right rotation to perform is:
 \begin{eqnarray}
  \label{eq:roty}
   R_x\left(p_0^{(f)},p_0^{(i)}\right)=e^{-\imath \frac{\theta_x}{2} \sigma_x}
 \end{eqnarray}
 with $\cos(\theta_x)=\left(1-2\; p_0^{(f)}\right)\left(1-2\; p_0^{(i)}\right)^{-1}$.
 For $b_y\neq 0$ we have solutions for $\cos(\theta_y)$ only if $-1/2\le1/2- p_0^{(f)}\le-\left(\left(1/2- p_0^{(i)}\right)^2+b_y^2\right)^{1/2}$.

\section{Conclusions}
\label{sec:conclusions}

We have worked out an exact dynamical
model with which we discussed the performance of a Quantum Otto
cycle in presence of {\it inner friction}. With respect to
previous approaches to the same problem, we obtained the growth of
polarization in the adiabatic branches of the cycle without any
ad-hoc assumption, but rather by following the dynamics generated
by the system Hamiltonian. In this way we have been able to deal
with the irreversibility of such transformations when they are
performed in a finite time and so to better characterize the whole
cycle. Finite time evolution leads to a decrease of values of
thermodynamical figures of merit for the heat engine, and we have
concentrated on the extractable work, the power and efficiency
which are partially quenched if inner friction is present.

The friction is related to the non-commutativity of the system and
control Hamiltonian, due to some misalignment between the internal
and control magnetic field axes. After explicitly studying its
effects for a fixed (and controlled) case, we turned to the more
realistic case in which such a misalignment is an unwanted
side-effect of the lack of control in the system, ultimately due
to the presence of disorder in the sample or inhomogeneity of the
magnetic field. The limiting cases are the completely ordered and
disordered samples; in the first one, we obtain an ideal quantum
Otto engine with efficiency given by
$\eta_{ideal}=1-\omega_1/\omega_2$; while for the completely
disordered case we showed that no positive work can be extracted
from the system, which cannot behave as an heat engine at all. In
the more general case of a finite-width Gaussian distribution of
tilting angles $\theta$, describing a given degree of disorder as
quantified by its variance, we obtained a quantitative description
of the efficiency reduction due to the disorder-induced inner
friction.

Finally, we proposed an optical experimental implementation of
such Otto cycle using the effective polarization qubit of a
photon, whose thermalization can be obtained by the coupling with
the spatial degree of freedom in a birefringent crystal.

\section{Acknowledgements}

We acknowledge funding from FIRB 2012 RBFR12NLNA, EU,  MICINN, CSIC,
FEDER funding under Grants FIS2011-23526 (TIQS),
FIS2014-60343-P (NOMAQ), postdoctoral JAE
program (ESF), FET-proactive project QuProCS and COST Action MP1209, and the UIB for Invited
Professors program.

\end{document}